\PassOptionsToPackage{dvipsnames}{xcolor}
\documentclass[twocolumn,tighten,twocolappendix]{aastex631}
\usepackage{gensymb}
\usepackage{natbib}      
\usepackage{graphicx}
\usepackage{amsmath}
\usepackage{times}
\usepackage{amssymb}
\usepackage{fancyhdr}
\usepackage{epsfig}
\usepackage{float}
\usepackage{url}
\usepackage{array}
\usepackage{verbatim}     
\usepackage{calc}
\usepackage{boldline}
\usepackage{tabu}
\usepackage{rotating}     
\usepackage{xcolor,colortbl}

\usepackage{multirow}

\usepackage{newtxtext,newtxmath}
\usepackage{aecompl}
\usepackage{ae,aecompl}
\usepackage{hyperref}
\usepackage{cleveref}
\usepackage{tabularx}
\usepackage{fontawesome}

\hypersetup{linkcolor=red,citecolor=MidnightBlue,filecolor=cyan,urlcolor=blue}
\newcommand{\p}[1]{{\color{magenta}{#1}}}
\newcommand{\q}[1]{{\color{red}{#1}}}

\begin{document}
%

\title{Optimization of Tessellation-based Statistics: Void Statistics}


\shorttitle{Optimized Void Statistics}
\shortauthors{Yu Liu et al.}

\author[0000-0002-9734-906X]{Yu Liu}
\affiliation{Institute of Physics, Laboratory of Astrophysics, \'Ecole Polytechnique F\'ed\'erale de Lausanne (EPFL), Observatoire de Sauverny, CH-1290 Versoix, Switzerland; y.liu@epfl.ch}
\affiliation{Department of Astronomy, Tsinghua University, Beijing, 100084, P.R. China}

\author[0000-0002-1991-7295]{Cheng Zhao}
\affiliation{Department of Astronomy, Tsinghua University, Beijing, 100084, P.R. China}

\author[0000-0002-9359-7170]{Yu Yu}
\affiliation{Department of Astronomy, School of Physics and Astronomy, Shanghai Jiao Tong University, Shanghai, 200240, P.R. China}
\affiliation{Key Laboratory for Particle Astrophysics and Cosmology (MOE)/Shanghai Key Laboratory for Particle Physics and Cosmology, P.R. China}

\author[0000-0002-3369-3718]{Zhejie Ding}
\affiliation{Nanjing Institute of Astronomical Optics and Technology, Chinese Academy of Sciences, Nanjing, Jiangsu, 210042, P. R. China}
\affiliation{University of Chinese Academy of Sciences, Nanjing 211135, P. R. China.}

\author[0000-0002-4616-4989]{Jean-Paul Kneib}
\affiliation{Institute of Physics, Laboratory of Astrophysics, \'Ecole Polytechnique F\'ed\'erale de Lausanne (EPFL), Observatoire de Sauverny, CH-1290 Versoix, Switzerland; y.liu@epfl.ch}
\textit{}

\begin{abstract}
Tessellation methods are extensively employed in the analyses of cosmic large-scale structure (LSS). However, these techniques are highly sensitive to perturbations in both densities and positions of points, often leading to substantial rearrangements of tessellation configurations. As a result, considerable additional statistical errors are introduced in various tessellation-based statistics, thereby weakening their cosmological constraints. In this work, we identify this issue and propose an efficacious measurement scheme through subsampling and averaging to enhance the stabilities of tessellation-based statistics. As a case study, we apply the new scheme to measure multiple primary void statistics [i.e., void size function (VSF), void two-point correlation function (VTCF), and void power spectrum (VPS)] in two distinct classes of voids, based on Delaunay and Voronoi tessellations, respectively. We notice that the statistical uncertainties in void statistics can be predominantly attributed to tessellation instabilities. Through rigorous testing, we demonstrate that the proposed method can substantially eliminate these scatters to deeply mine the statistical power of void statistics. Specifically, we find that our method can dramatically boost the signal-to-noise ratios (SNRs) of void Baryon Acoustic Oscillations (BAOs) and significantly improve the constraining power of void statistics on cosmological parameters. These findings showcase enormous application potentials of our new method in maximizing extraction of cosmological information from galaxy surveys. Importantly, our method is simple yet highly potent with broad applicability, hopefully evolving into a standard framework for measuring tessellation-based statistics in the future.
\end{abstract}


\section{Introduction} \label{sec:introduction}
Galaxy surveys are stepping into an exciting Stage-IV era (\citealt{2006astro.ph..9591A}), with a series of cutting-edge experiments in operation or on the brink of commencing [e.g., SPHEREX (\citealt{2014arXiv1412.4872D}), PFS (\citealt{2014PASJ...66R...1T}), Roman (\citealt{2015arXiv150303757S}), DESI (\citealt{2016arXiv161100036D}), LSST (\citealt{2019ApJ...873..111I}), 4MOST (\citealt{2019Msngr.175....3D}), CSST (\citealt{2019ApJ...883..203G}), HETDEX (\citealt{2021ApJ...923..217G}), Euclid (\citealt{2024arXiv240513491E}), etc.]. The ambitious so-called "Stage-V" surveys [e.g., MegaMapper (\citealt{2019BAAS...51g.229S}), MSE (\citealt{2019arXiv190303158P}), WST (\citealt{2024arXiv240305398M}), MUST (\citealt{2024arXiv241107970Z}), ESST (\citealt{2024SCPMA..6779511S}), Spec-S5 (\citealt{2025arXiv250307923B}), etc.] are also being successively proposed, with survey efficiency and information content exceeding those of the current state-of-the-art DESI by over tenfold (\citealt{2022arXiv220903585S, 2024arXiv241107970Z, 2025arXiv250307923B, 2025SCPMA..6880403C}). Tantalizingly, all these endeavors will bring great opportunities for revolutionizing our understanding of various major questions in fundamental physics (e.g., inflation model, dark energy, gravity property, neutrino mass, dark matter, etc.; cf.\ \citealt{2022arXiv221109978C}), heralding a golden age for cosmology.

Extraction of cosmological information from these massive surveys relies on effective statistical methods. Historically, traditional two-point statistics have long dominated the LSS analyses (\citealt{1980lssu.book.....P, 2003moco.book.....D, 2004ApJ...606..702T}). These methods provide a sufficient characterization of Gaussian or mildly non-Gaussian density fluctuations in early Universe and have played a pivotal role in constraining cosmological models (\citealt{2013ApJS..208...19H, 2020A&A...641A...6P}). However, their reliance on second-order statistics inherently limits their sensitivities to complex non-Gaussian features emerging from LSS non-linear structure formation in late Universe (\citealt{2002PhR...367....1B, 2024arXiv240502252B}). To capture complementary higher-order information, additional non-Gaussian statistics are required, e.g., three/four-point statistics (\citealt{2003MNRAS.340..580T, 2006PhRvD..74b3522S, 2015JCAP...05..007B, 2021arXiv210801670P, 2022MNRAS.509.2457P}), Minkowski functionals (\citealt{1994A&A...288..697M, 1997ApJ...482L...1S, 2020PhRvD.101f3515L, 2021PhRvD.104j3522M, 2024ApJS..273...33L}), scattering transform (\citealt{2020MNRAS.499.5902C, 2022PhRvD.106j3509V, 2022PhRvD.105j3534V, 2024JCAP...11..061V, 2024PhRvD.109j3503V}), density-split clustering (\citealt{2021MNRAS.505.5731P, 2023MNRAS.522..606P, 2024MNRAS.531..898P, 2024MNRAS.531.3336C, 2025JCAP...01..026M}), void statistics (\citealt{2015A&C.....9....1S, 2016MNRAS.459.2670Z, 2025arXiv250322532C, 2025A&A...695A..19F}), etc.


In particular, void statistics, quantifying the abundance, morphology, clustering, and other aspects of under-dense regions, are actively employed in various LSS studies, e.g., neutrino mass (\citealt{2015JCAP...11..018M, 2019MNRAS.488.4413K, 2021MNRAS.504.5021C, 2023JCAP...08..010V, 2024ApJ...969...89T}), modified gravity (\citealt{2012MNRAS.421.3481L, 2015MNRAS.451.1036C, 2018MNRAS.475.3262F, 2019A&A...632A..52P, 2018PhRvD..98b3511B}), dark energy (\citealt{2010PhRvD..82b3002B, 2012MNRAS.426..440B, 2015PhRvD..92h3531P, 2019JCAP...12..040V, 2022A&A...667A.162C}), Alcock-Paczynski effect (\citealt{2012ApJ...761..187S, 2017ApJ...835..160M, 2020MNRAS.499..587E, 2022A&A...658A..20H, 2024A&A...691A..39R}), baryon acoustic oscillation (BAO) (\citealt{2016PhRvL.116q1301K, 2016MNRAS.459.4020L, 2020MNRAS.491.4554Z, 2022MNRAS.511.5492Z, 2023MNRAS.526.2889T}), redshift space distortion (\citealt{2016MNRAS.462.2465C, 2017PhRvD..95f3528C, 2017JCAP...07..014H, 2021MNRAS.500..911C, 2022MNRAS.509.1871C}), integrated Sachs-Wolfe effect (\citealt{2013A&A...556A..51I, 2014ApJ...786..110C, 2017MNRAS.466.3364C, 2018MNRAS.475.1777K, 2021MNRAS.500.3838D}), cosmological constraints (\citealt{2019PhRvD.100l3513A, 2020JCAP...12..023H, 2022MNRAS.513..186A, 2023ApJ...953...46C, 2025A&A...695A..19F}), etc. Since voids encode unique information about gravitational collapse (\citealt{2011IJMPS...1...41V, 2016PhRvL.117i1302H}), their combination with other summary statistics can help break degeneracies and significantly tighten cosmological constraints (\citealt{2016ApJ...820L...7S, 2019PhRvD..99f3525S, 2021ApJ...919...24B, 2022MNRAS.511.5492Z, 2024ApJ...976..244S}).

There exists diverse methodologies to delineate void domains with distinct constraints tailored for specific research goals (\citealt{2008MNRAS.387..933C, 2018MNRAS.476.3195C, 2025arXiv250322532C}). Currently, the mainstream void-finding tools are mostly based on tessellation algorithms, i.e.,  Delaunay tessellation [e.g., WVF (\citealt{2007MNRAS.380..551P}), DIVE (\citealt{2016MNRAS.459.2670Z}), Delfin++ (\citealt{2023A&C....4400713G}), etc.\ (\citealt{2018A&C....22...48A})] and Voronoi tessellation [e.g., ZOBOV (\citealt{2008MNRAS.386.2101N}), VIDE (\citealt{2015A&C.....9....1S}), REVOLVER (\citealt{2019PhRvD.100b3504N}), VAST (\citealt{2022JOSS....7.4033D}), VEGA (\citealt{2025arXiv250116431G}), etc.]. These spatially self-adaptive methods can capture multi-scale geometry of local point distribution, which helps outline the presence and shape of void-like regions (\citealt{2009LNP...665..291V}). However, these techniques exhibit high sensitivities to perturbations in point number density or position, always resulting in substantial rearrangements of tessellations, due to their strict spatial partitioning restrictions (\citealt{2009LNP...665..291V, 2021MNRAS.503..557A}). This instability issue will be inherited by the tessellation-based void identifications and inevitably leads to additional statistical uncertainties in corresponding void statistics. As a result, it considerably weakens the constraining power of various tessellation-based void statistics.

In this study, for the first time, we notice this critical issue and address it with a new measurement scheme for void statistics, relying on subsampling and averaging. To validate the effectiveness of our method, two representative void finders, one based on Delaunay tessellation [i.e., DIVE\footnote{\url{https://github.com/cheng-zhao/DIVE}} (\citealt{2016MNRAS.459.2670Z})] and the other on Voronoi tessellation [i.e., VIDE\footnote{\url{https://bitbucket.org/cosmicvoids/vide_public/src/master/}} (\citealt{2015A&C.....9....1S})], are employed in our tests. Our analyses indicate that, in most cases, especially for sparse sample scenarios, these statistical uncertainties are primarily induced by tessellations. We reveal that our method can significantly suppress these scatters on various major void statistics, including VSF, VTCF, and VPS (see Section~\ref{sec:statistics}). In our testing domain, we observe an enhancement of their cosmological constraining power by $\sim20\text{--}40\%$, through utilization of our new method. We further demonstrate that, for large voids identified by DIVE, in the absence of shot noises, their void BAO signatures can closely resemble those of linear BAOs. Moreover, by employing our new method, these voids exhibit a large increase in void BAO SNRs, even surpassing that of halos. Given the current landscape (\citealt{2026arXiv260114362C}), our work is well-timed and offers an improved and more robust approach for extracting cosmological information from LSS using void statistics in future galaxy surveys.

This paper is structured as follows. Section~\ref{sec:tessellations} provides a concise overview of Delaunay and Voronoi tessellations. Section~\ref{sec:voids} and Section~\ref{sec:optimization} introduce the adopted tessellation-based void finders and our optimized measurement scheme for tessellation-based statistics. In Section~\ref{sec:data}, we detail the cosmological $N$-body simulations and halo catalogues used in this study. Then, in Section~\ref{sec:statistics}, we present comprehensive and well-supported results for VSF, VTCF, and VPS by comparing measurements from our optimized scheme against traditional scheme. Finally, we give summaries and discussions in Section~\ref{sec:summary}. In addition, Appendix~\ref{app:subsampling} provides a proof for our bootstrap subsampling method. Appendix~\ref{app:Field_Estimator} provides a brief introduction to two representative tessellation-based density field estimators. Appendix~\ref{app:VPS_bias_SN} displays the results for VPSs, shot noises, and biases of all DT voids and disjoint DT voids. Appendix~\ref{app:Fisher_forecast} describes certain technical aspects for the Fisher error forecasts employed in this study.

\section{Delaunay and Voronoi Tessellations} \label{sec:tessellations}
\begin{figure*}
\centering 
\includegraphics[width=0.95\textwidth]{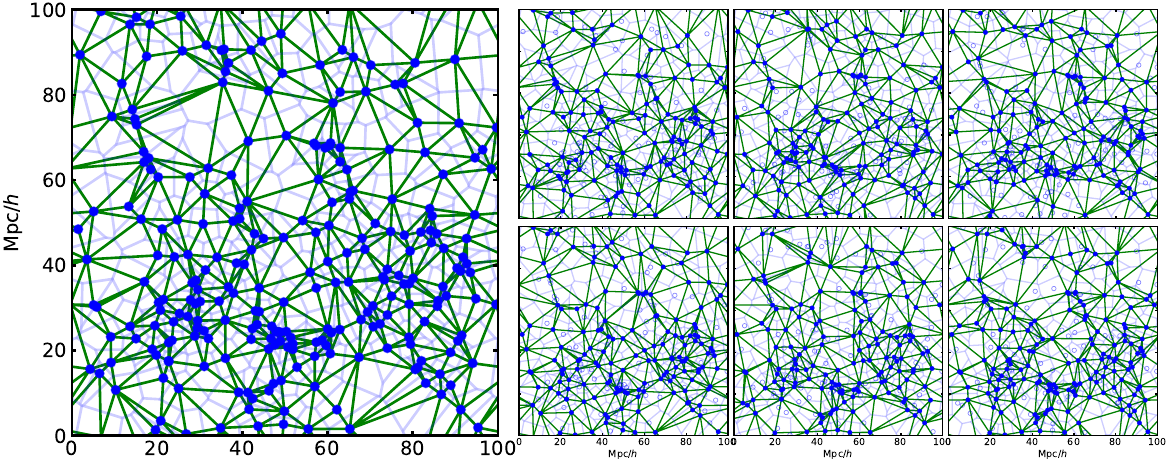}
\caption{Delaunay tessellations (green lines) and Voronoi tessellations (pale-blue lines) constructed from point sets (blue dots) in a 2-dimensional square box of width $100\,h^{-1}\mathrm{Mpc}$ with periodic boundary conditions. Here, these blue dots mark the two‑dimensional projected positions of halos within a random cubic volume of $(100\,h^{-1}\mathrm{Mpc})^3$ from one of our cosmological $N$-body simulations, with a number density of $\bar{n}_\mathrm{h} = 2.28 \times 10^{-4}\,(h^{-1} \mathrm{Mpc})^{-3}$ (see Section~\ref{sec:data}). The blue dots in the six small subpanels on the right are subsamples of the point set in the large panel on the left, obtained via bootstrap subsampling (see Section~\ref{sec:optimization}), with open circles denoting the positions of points not selected during bootstrap subsampling. While the six right-hand subpanels contain roughly the same number of points, the positions of some points differ. Contrasting the large panel with the six smaller ones highlights how point number density significantly alters the tessellation configurations. The comparisons among the six right-hand subpanels reveal that variations in point positions also lead to remarkable rearrangements in the resulting tessellations, despite these six subsamples originate from the same original dataset. These facts demonstrate that tessellations are highly sensitive to perturbations in point distributions, leading to tessellation instabilities. As a consequence, the statistics built upon tessellations are also susceptible to these instabilities, manifesting as additional statistical uncertainties.}
\label{fig:tessellations}
\end{figure*}

Delaunay and Voronoi tessellations are basic concepts in stochastic and computational geometry (\citealt{StoyanKendalMecke+1987, 10.5555/261226}). These methods can elegantly partition space into a union of space-filling and mutually disjoint cells based on a point distribution, uniquely and self-adaptively. Given a countable set $\mathcal{P}$ of points $\left\{\mathbf{x}_{1}, \ldots, \mathbf{x}_{N} \right\}$ in $D$-dimensional space, Delaunay tessellation imposes the circumsphere $\mathcal{S}_{i}$ of each Delaunay cell $\mathcal{D}_{i}$ does not contain any other points in $\mathcal{P}$ (\citealt{Delaunay_1934aa, 2009LNP...665..291V})
\begin{eqnarray}\label{eq:1}
\mathcal{D}_{i} = \mathcal{T}\left(\mathbf{x}_{i1}, \ldots, \mathbf{x}_{i(D+1)}\right) \ \text{with} \ d\left(\mathbf{C}_{i}, \mathbf{x}_{j}\right) > R_{i} && \nonumber\\
\forall j \neq i1, \ldots, i(D+1),
\end{eqnarray}
where $R_{i}$ and $\mathbf{C}_{i}$ represent the radius and circumcenter of $\mathcal{S}_{i}$ respectively, $\mathcal{D}_{i}$ is the simplex $\mathcal{T}$ [i.e., triangle (tetrahedron) in $2$($3$)-dimensional space] defined by $D+1$ points $\left\{\mathbf{x}_{i1}, \ldots, \mathbf{x}_{i(D+1)} \right\} \in \mathcal{P}$. Correspondingly, Voronoi tessellation assigns each point $\mathbf{x}_{i} \in \mathcal{P}$ a Voronoi cell $\mathcal{V}_i$ [i.e., Polygon (Polyhedron) in $2$($3$)-dimensional space], any point in which is nearer to $\mathbf{x}_{i}$ than to any other points in $\mathcal{P}$ (\citealt{Voronoi1908, 2000stca.conf.....O}):
\begin{equation}\label{eq:2}
\mathcal{V}_{i} = \left\{\mathbf{x} \mid d\left(\mathbf{x}, \mathbf{x}_{i}\right) < d\left(\mathbf{x}, \mathbf{x}_{j}\right)\right\} \quad \forall j \neq i.
\end{equation}
Two tessellations are each other's dual, allowing for direct inference from one to the other and vice versa, e.g., the circumsphere centre of a Delaunay cell is actually a vertex of a Voronoi cell (see Figure~\ref{fig:tessellations}).

Tessellation methods can self-adaptively adjust their cell sizes and shapes to delineate the hierarchical and anisotropic characteristics of spatial point distribution, making them play a key role in various LSS algorithms, e.g., cosmological hydrodynamical simulation (\citealt{2010MNRAS.401..791S, 2020ApJS..248...32W}), structure classification (\citealt{2007A&A...474..315A, 2010ApJ...723..364A, 2013MNRAS.429.1286C}), and void identification (\citealt{2008MNRAS.386.2101N, 2015A&C.....9....1S, 2016MNRAS.459.2670Z}; see Section~\ref{sec:voids}), etc. In cosmology context, such tessellation-based algorithms commonly operate on the basis of reconstructing underlying density fields from discrete tracers, e.g., particles/halos in numerical simulations and galaxies/quasars in observational surveys (\citealt{2000A&A...363L..29S, 2009LNP...665..291V, 2021MNRAS.503..557A, 2025MNRAS.536..807F}). Representative field reconstruction methods are typified by the zeroth-order Voronoi Tessellation Field Estimator (VTFE) and the first-order Delaunay Tessellation Field Estimator (DTFE). The noise characteristics of reconstructed fields can be attributed to two main sources: natural sampling noise arising from the finite number of discrete tracers, which inevitably leads to information loss, and additional contribution from tessellation-induced artifacts, associated with the intrinsic tessellation kernels determined by the local density and geometry of point process (cf.\ \citealt{2009LNP...665..291V}) (see Appendix~\ref{app:Field_Estimator} for further details). 

Meanwhile, tessellation configurations are intrinsically sensitive to point positions (see \citealt{Goberna2011OnTS, doi:10.1142/S0218195913600078, doi:10.1142/S021819591450006X, 2021MNRAS.503..557A}) and number densities, as a result of their extremely stringent definition constraints (see Equation~\ref{eq:1} and Equation~\ref{eq:2}). This can be visually evident through a comparison of different panels in Figure~\ref{fig:tessellations}, where small perturbations in point distributions can lead to substantial rearrangements in tessellation configurations\footnote{In practice, perturbations in tracer samples are ubiquitous, arising from various sources of numerical and observational uncertainties, such as differences in simulation resolution, halo-finding algorithms, galaxy redshift uncertainties, and sample selection effects, etc.}. The sensitivities are also expected to propagate into tessellation-based statistics, leading to statistical instabilities, manifested as additional statistical errors (see Section~\ref{sec:statistics}). For instance, different subsamplings of halos (see the six small subpanels in Figure~\ref{fig:tessellations}) can substantially reshuffle tessellation structures, with particularly strong implications for void identification using DIVE (see Section~\ref{sec:voids}), where the inferred void populations can differ significantly between realizations. Naturally, it also introduces uncertainties in the positions, as well as the shapes and volumes, of VIDE-identified voids (see Section~\ref{sec:voids} and Section~\ref{sec:statistics}). Therefore, this might be more intuitive to recognize that the properties (including size, clumpiness, etc.) of tessellation-defined objects inherently carry additional tessellation-induced statistical uncertainties, beyond the intrinsic cosmic variance.

Extra artificial smoothing procedures [e.g., Gaussian smoothing (\citealt{2007A&A...474..315A, 2010ApJ...723..364A, 2013MNRAS.429.1286C}) and Natural Neighbour Rank filtering (\citealt{2007MNRAS.380..551P})] can attenuate discreteness effects, but inevitably at the cost of erasing substantial yet intrinsically subtle structures (\citealt{2024ApJS..273...33L}). In an attempt to ameliorate the DTFE artifacts, efforts have been made to replace the linear interpolation with higher-order, more sophisticated methods such as Natural Neighbour interpolation and Kriging interpolation. However, tests indicate that these methods do not appear to provide any improvements for void identification (\citealt{2011MNRAS.416.2494P}). A more recent related development is the Phase-Space DTFE (PS-DTFE) formalism (\citealt{2025MNRAS.536..807F}), which combines phase-space density estimation (\citealt{2012PhRvD..85h3005S, 2012MNRAS.427...61A}) with DTFE. While it effectively mitigates artifacts in multi-stream regions, the reconstructed field exhibits discontinuities at fold caustics. Moreover, it relies on prior knowledge of tracers before shell crossing, which limits its applicability to particle and halo distributions derived from $N$-body simulations. In principle, such crafted refinements built upon a single-step tessellation mesh can hardly escape from the tessellation sensitivities (i.e., the tendency of tessellation to fluctuate), and thus the resulting density fields remain subject to tessellation-induced uncertainties.

\begin{figure*}
\centering 
\includegraphics[width=0.851\textwidth]{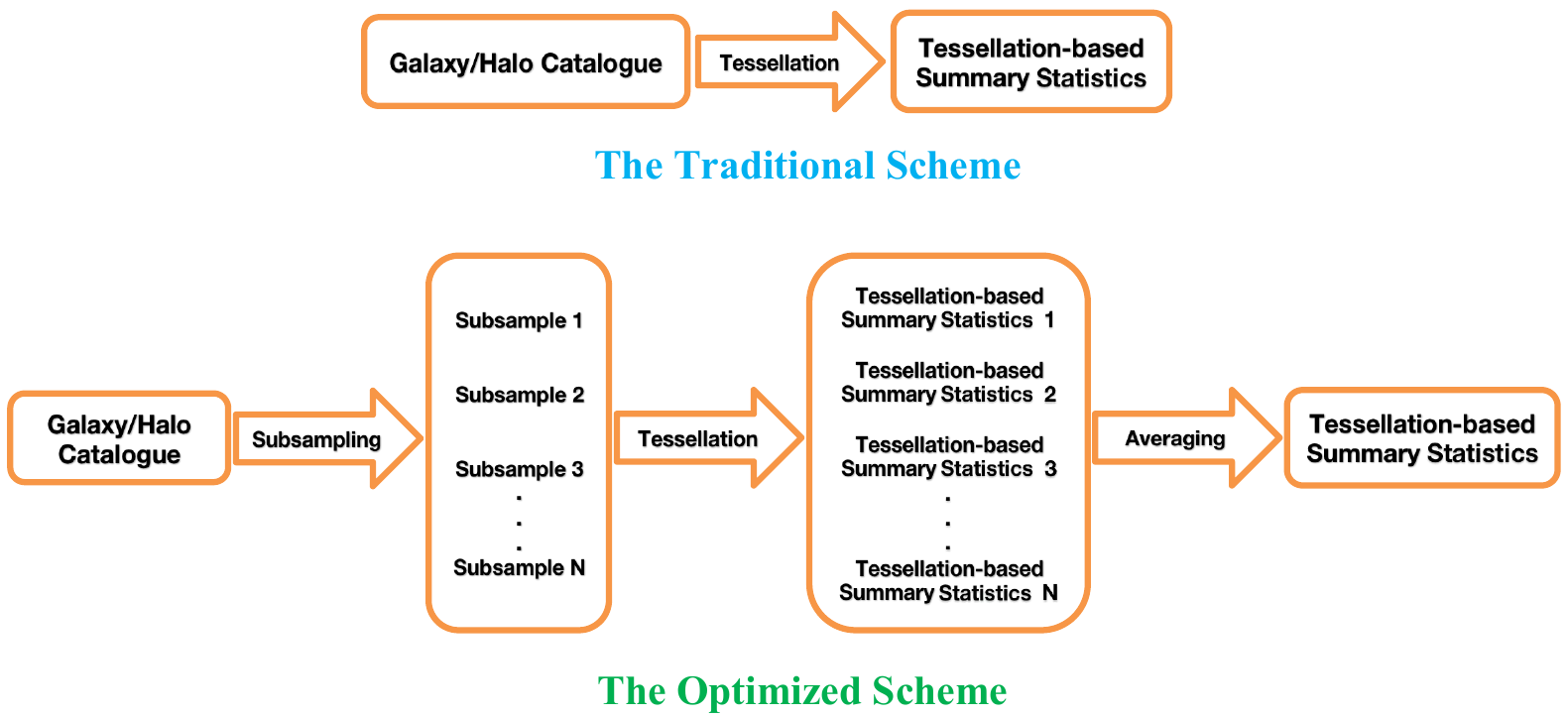}
\caption{Flowchart of traditional and optimized schemes for measuring tessellation-based statistics. In traditional scheme, tessellation-based statistics are computed directly from the tessellation constructed from original halo/galaxy catalogue. In contrast, the optimized scheme we advocate involves generating multiple subsamples through downsampling of original halo/galaxy catalogue. For each subsample, a separate tessellation is constructed and the corresponding tessellation-based statistics are subsequently calculated. These statistics are ultimately averaged across all subsamples to obtain the final measurements. In this work, as a case study, we focus on void statistics as representative tessellation-based statistics. Our proposed optimization method can significantly improve the statistical stabilities of tessellation-based statistics, reduce statistical uncertainties, and enhance their constraining power on cosmological parameters (see Section~\ref{sec:statistics}).}
\label{fig:flow_char}
\end{figure*}

A further effective stochastic DTFE approach is to generate an ensemble of tracer realizations via constrained point jittering, compute the DTFE field for each realization, and then take the ensemble average. The mean field can suppress the sharp geometric artifacts arising from deterministic Delaunay tessellation in individual realizations while preserving key DTFE properties such as anisotropy, self-scaling, and mass conservation. This method shares a similar “averaging” philosophy with the optimized scheme advocated in this paper (see Section~\ref{sec:optimization}), suggesting that it may also substantially mitigate the tessellation-induced sensitivities and thereby reduce the associated statistical errors\footnote{Nevertheless, further detailed tests are required before any definitive conclusion can be drawn. Moreover, as its construction of position-perturbed samples alters the original data, whether and to what extent this may introduce biases in cosmological inference remains unclear.}. However, a key difference lies in the level at which averaging is performed: their approach operates at the field level, whereas ours is carried out at the level of statistics (see Figure~\ref{fig:flow_char} and Section~\ref{sec:optimization}). Naturally, our framework can also be extended to perform field-level averaging to assess its effectiveness in suppressing DTFE artifacts, which we leave for future work. If such a mean field indeed can provide a more stable basis for defining LSS structures, it may in turn lead to more robust summary statistics of the identified structures. If feasible, this would reduce statistical estimation to a single computation, avoiding the need to evaluate statistics across multiple subsamples and thereby significantly lowering the computational cost.

\begin{figure*}
\centering 
\includegraphics[width=0.813\textwidth]{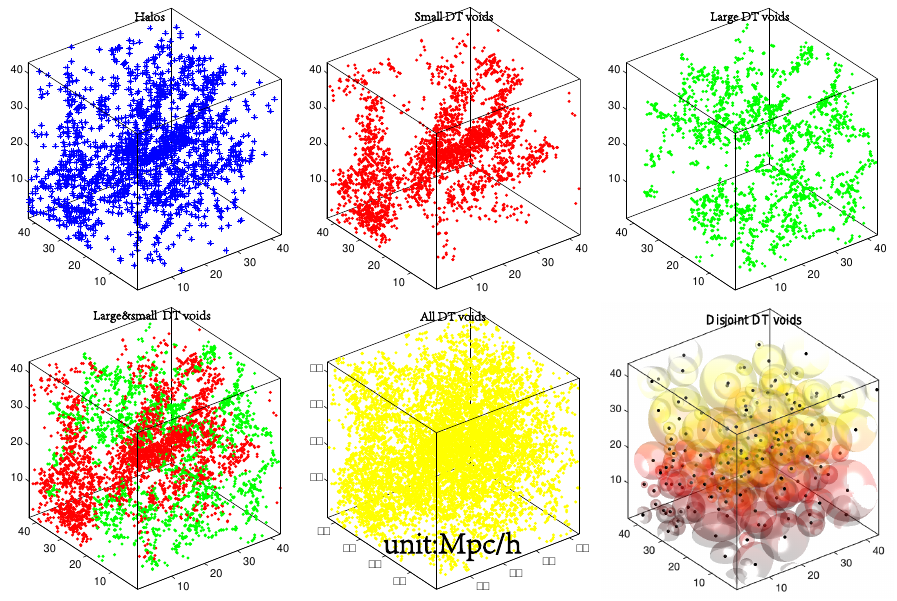}
\caption{Three-dimensional spatial distributions of halos and various types of DT voids in a cubic box with a side length of $43.5\,h^{-1}\mathrm{Mpc}$. In the top panels (from left to right), we show the positions of halos, small DT voids, and large DT voids, marked by blue asterisks, red dots, and green dots, respectively. The bottom panels display (from left to right) the combined spatial distribution of large\&small DT voids, all DT voids (yellow dots), and the corresponding disjoint DT voids (black dots). In the bottom-right panel, the disjoint DT voids are visualized as semitransparent spheres. As shown, small DT voids follow a spatial distribution similar to that of halos, preferentially tracing high-density regions, while large DT voids tend to populate low-density regions, complementing the small ones. Due to the non-overlapping constraint (see Section~\ref{sec:voids}), the number density of disjoint DT voids is relatively low. In addition, two-dimensional spatial comparisons between DT voids and their corresponding halos/galaxies can be found in the existing literature (e.g., \citealt{2016MNRAS.459.2670Z, 2016PhRvL.116q1301K, 2016MNRAS.459.4020L, 2020MNRAS.491.4554Z, 2023MNRAS.526.2889T}).}
\label{fig:DT_voids_halos}
\end{figure*}
\section{Tessellation-based Void Finders} \label{sec:voids} 

Cosmic voids are generic low-density regions of the Universe. Despite they are among the most distinctive and striking features of the cosmic web, there exists no unequivocal void definition in LSS studies due to persistent disagreements about their precise boundaries. Thus, a plethora of void-finding methods designed for different research interests are available on the market, with the identified voids exhibiting diverse geometric outlines, e.g., spherical (e.g., \citealt{2005MNRAS.363..977P, 2016MNRAS.459.2670Z}) or arbitrary (e.g., \citealt{2007MNRAS.380..551P, 2008MNRAS.386.2101N}) shapes. In this work, we opt for two representative parameter-free void finders, i.e., DIVE (\citealt{2016MNRAS.459.2670Z}) and VIDE (\citealt{2015A&C.....9....1S}), which are based on Delaunay and Voronoi tessellations, respectively, to validate our optimization approach.

DIVE employs publicly available Computational Geometry Algorithms Library (CGAL\footnote{\url{https://www.cgal.org/index.html}}) to implement Delaunay tessellation based on spatial distribution of halos/galaxies. The algorithm simply designates the circumspheres of Delaunay tetrahedra (see Section~\ref{sec:tessellations}) as Delaunay Triangulation (DT) voids, and the centers and radii of these circumspheres are straightforwardly calculated as outputs. Diverging from conventional void notions, DT voids exhibit severe overlap, with notably high number density (i.e., low shot-noise level) around seven times that of the corresponding halo/galaxy sample (see \citealt{2016MNRAS.459.2670Z} and Figure~\ref{fig:DT_voids_halos}), rendering them as tracers exceptionally suitable for clustering analyses, e.g., measuring BAO signatures (\citealt{2016PhRvL.116q1301K, 2016MNRAS.459.4020L, 2022MNRAS.511.5492Z}) and more. DT voids can be further categorized into two distinct classes according to their sizes. Among them, small DT voids are treated as voids-in-clouds which trend to reside in contracting high-density regions, while large DT voids are considered to be voids-in-voids which are likely to trace expanding low-density regions (\citealt{2004MNRAS.350..517S}). Moreover, based on DT void sample, one can also construct the so-called "disjoint" (or "tunnel") void catalogue by discarding the DT voids that overlap with larger voids (or the DT voids whose centers inside larger voids), which helps avoid identifying the same regions as voids multiple times (see Figure~\ref{fig:DT_voids_halos} and \citealt{2016MNRAS.459.2670Z, 2018MNRAS.476.3195C}).

VIDE is built upon ZOBOV algorithm (\citealt{2008MNRAS.386.2101N}) as its core to identify voids irrespective of sizes and shapes. It offers flexible functionality for handling halos in periodic simulation boxes or survey galaxies entangled with selection functions and masks. Given a halo/galaxy catalogue, the algorithm estimates density field with VTFE technique (see Section~\ref{sec:tessellations} and Section~\ref{app:Field_Estimator}), which assigns constant density across each Voronoi cell based on the inverse of its volume. Via watershed transform, the Voronoi cells surrounding local density minima are grouped into extended "catchment basin" zones as the recognized voids. For building void hierarchy, the zones considered as sub-voids can be merged to assemble larger agglomerations treated as parent-voids. Within ZOBOV, the merged zones share the common separating ridges below $20\%$ of the mean tracer density (\citealt{2008MNRAS.386.2101N, 2015A&C.....9....1S}), which helps reduce the misclassification of Poisson fluctuations as voids and prevents the entire volume from being identified as a void. Actually, in such a setup, merging occurrences are typically infrequent. In this work, for constructing samples of voids without any sub-voids, the merging threshold of a default minimal value of $10^{-9}$ is established to prevent the merging of adjacent zones. Ultimately, as a versatile tool for comprehensive void analyses, VIDE calculates multiple key void quantities, e.g., barycenters, ellipticities, and effective radii, as outputs.

According to the above description, throughout this paper, we tend to refer to ZOBOV/VIDE rather than VIDE alone, as ZOBOV constitutes the tessellation and watershed core of the pipeline. Moreover, a fundamental distinction exists between DIVE and ZOBOV/VIDE [as well as other watershed-based tools, e.g., WVF (\citealt{2007MNRAS.380..551P}), V$^2$ (\citealt{2022JOSS....7.4033D}), and REVOLVER (\citealt{2019PhRvD.100b3504N})]: DIVE directly utilizes tessellation to define voids, whereas the latter relies on tessellation-based density field reconstruction (cf.\ Appendix~\ref{app:Field_Estimator}) followed by a separate watershed-based void-finding procedure. Here, the two steps in ZOBOV/VIDE are clearly distinct from each other: the tessellation analysis is a purely geometric, local instrument, whereas the watershed void finding is a topological formalism that seeks to identify void boundaries via locations of saddle points in the reconstructed density fields.

\section{Optimized Measurement Scheme for Tessellation-based Statistics} \label{sec:optimization}
In previous traditional studies, tessellations are directly performed based on a given halo/galaxy catalogue, and tessellation-based statistics (i.e., void statistics as a case study in this paper, with other tessellation-based LSS statistics left for future work) are subsequently measured as the final results (see Figure~\ref{fig:flow_char}). However, due to the intrinsic sensitivities (or instabilities) of tessellation methods (see Section~\ref{sec:tessellations}), substantial additional statistical uncertainties are typically introduced in the measured statistics, degrading their cosmological constraining power (see Section~\ref{sec:statistics}). To address or alleviate this critical issue, we propose a solution-oriented measurement scheme by leveraging a strategy of subsampling and averaging. The procedure is as follows: (i) given a halo/galaxy catalogue, random subsamplings are performed to generate $N$ halo/galaxy subsamples; (ii) for each subsample, tessellation is constructed individually, followed by the measurement of tessellation-based statistics; (iii) the statistics derived from measurements of different subsamples are averaged to yield the final results (see Figure~\ref{fig:flow_char}). In Section~\ref{sec:statistics}, we will demonstrate that our new method can effectively immunize tessellation-based statistics against tessellation instabilities, markedly suppressing statistical uncertainties.

In practice, the choice of subsampling methods is somewhat arbitrary. In this work, we employ the bootstrap method, a resampling technique initially introduced in \citealt{e89fac9c-03d7-3e22-aa30-08f5596f8fce} and commonly used in statistics. The core of this method lies in resampling the original sample (i.e., halo/galaxy catalogue) with replacement, ensuring that the number of resampling iterations exactly matches the sample size of the original catalogue (i.e., the total number of halos/galaxies). This means that each halo/galaxy in the original sample can be sampled multiple times during the procedure. For the cases of large sample size, approximately $1 - 1/e \simeq 0.63212$ of the halos/galaxies in the original catalogue are sampled at least once, forming an acquired subsample, where Euler's number $e \simeq 2.71828$ (see Appendix~\ref{app:subsampling}). Here, note that each halo selected at least once is included only once in a given subsample, i.e., no duplicate halos are present. To adequately cover the original sample, the number of subsamples should satisfy $N > \ln{n_{\mathrm{h}}}$\footnote{Alternative subsampling methods can be devised to circumvent this limitation. For instance, randomly shuffling the original halo/galaxy catalogue and partitioning it into overlapping subsamples, with shared halos/galaxies between adjacent subsamples. This patching strategy can guarantee full dataset coverage while preserving local representativeness within each subsample. As designing an optimal subsampling approach is not the concern of this work, we leave it for future investigation.}, where $n_{\text{h}}$ is the total number of halos/galaxies. Notably, the subsamples constructed in this manner not only partially differ in halo/galaxy positions, but also exhibit fluctuation in sample size $n_{\rm{sub}}$, which follows a binomial distribution $n_{\mathrm{sub}} \sim \mathrm{B}\left(n_{\mathrm{h}}, 1-e^{-1}\right)$.

In our tests for the new measurement scheme (see Section~\ref{sec:statistics}), we produce $N=100$ and $N=30$ subsamples for DT and ZOBOV/VIDE void scenarios, respectively, from each original halo catalogue. Here, these extensive sets of subsamples are intended to facilitate subsequent, more detailed convergence testing in our following studies. Indeed, these subsample numbers are sufficiently large to ensure adequate coverage of the original halo catalogue and to guarantee convergences of the results. For instance, each of the densest halo catalogues considered in this work (see Section~\ref{sec:data}) contains $n_\mathrm{h}=216\times10^{5}$ halos, yielding $\ln n_\mathrm{h} \simeq 17$. Therefore, while subsampling framework renders each individual subsample sparser, the adopted new scheme does not discard any of the original data; rather, it simply alters the way the data are processed.

\begin{figure*}
\centering
\includegraphics[width=1.0\textwidth]{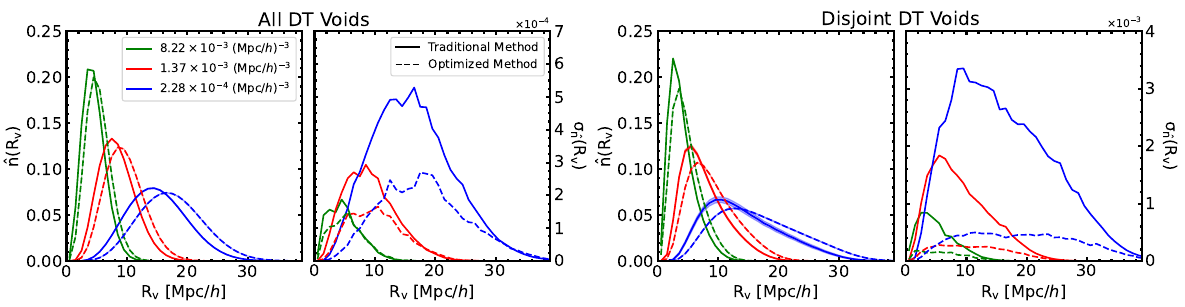}
\caption{VSFs (left subpanels) and their corresponding $1$-$\sigma$ uncertainties (right subpanels) for all DT voids (left panel) and disjoint DT voids (right panel). Green, red, and blue lines correspond to halo catalogues with number densities of $\bar{n}_\mathrm{h} = 8.22 \times 10^{-3} \, (h^{-1} \mathrm{Mpc})^{-3}$, $\bar{n}_\mathrm{h} = 1.37 \times 10^{-3} \, (h^{-1} \mathrm{Mpc})^{-3}$, and $\bar{n}_\mathrm{h} = 2.28 \times 10^{-4} \, (h^{-1} \mathrm{Mpc})^{-3}$, respectively (the same below). In the left subpanel of each panel, each curve and its shaded region represent the mean and standard deviation (see the right subpanel of each panel) of the VSFs obtained from 40 simulation realizations with BAO features, except for the highest-density case, where 10 realizations are used (see Section~\ref{sec:data}; the same below). Solid and dashed lines correspond to the results measured with traditional and optimized schemes, respectively (the same below). As demonstrated, optimized scheme yields significantly smaller scatters than traditional scheme, because subsampling-and-averaging strategy helps stabilize the measurements of tessellation-based statistics (see Section~\ref{sec:optimization}).}
\label{fig:DT_all_disjoint_VSF}
\end{figure*}

\section{Cosmological $N$-body Simulations and Data Samples} \label{sec:data}
We employ $40$ pairs of cosmological $N$-body simulations realized by {\tt FastPM} (\citealt{2016MNRAS.463.2273F}). They adopt Planck flat $\Lambda$CDM cosmology parameterized with $[\Omega_\mathrm{c}$, $\Omega_\mathrm{b}$, $h$, $n_\mathrm{s}$, $\sigma_8]$ = $[0.2589$, $0.0486$, $0.6774$, $0.9667$, $0.8159]$ (\citealt{2016A&A...594A..13P}). Each simulation evolves $N_\mathrm{p} = 2048^3$ particles with a mass of $2.611\times10^{10}\,h^{-1}\mathrm{M}_\odot$ in a periodic cubic box of width $L_{\mathrm{box}} = 1380\,h^{-1}\mathrm{Mpc}$. To cancel cosmic variance in BAO analyses, simulation pair shares the same white-noise realization for generating initial conditions, with one including BAO features and the other excluding them. Dark matter halos are identified using friends-of-friends halo finding module implanted in {\tt nbodykit}\footnote{\url{https://nbodykit.readthedocs.io/en/latest/index.html}} (\citealt{2018AJ....156..160H}), where linking length is set to $b = 0.2$.

To investigate the impacts of halo number densities on measurements of void statistics, three sets of samples at $z = 0.0$ are compiled, consisting of 10 pairs, 40 pairs, and 40 pairs of halo catalogues with corresponding number densities of $\bar{n}_\mathrm{h} = 8.22 \times 10^{-3} \, (h^{-1} \mathrm{Mpc})^{-3}$, $\bar{n}_\mathrm{h} = 1.37 \times 10^{-3} \, (h^{-1} \mathrm{Mpc})^{-3}$, and $\bar{n}_\mathrm{h} = 2.28 \times 10^{-4} \, (h^{-1} \mathrm{Mpc})^{-3}$, respectively. These catalogues are constructed by discarding halos with masses below cutoffs of $M_{\mathrm{min}} \simeq 4.18 \times 10^{11} \, h^{-1} \mathrm{M}_\odot$, $M_{\mathrm{min}} \simeq  3.42 \times 10^{12} \, h^{-1} \mathrm{M}_\odot$, and $M_{\mathrm{min}} \simeq 2.15 \times 10^{13} \, h^{-1} \mathrm{M}_\odot$, respectively. Our sample selections are motivated by the fact that galaxy catalogues in most observational surveys are constrained by faint flux limits (or low mass limits). Throughout this paper, subhalos are excluded from analyses.

\section{Measurements of Void Statistics} \label{sec:statistics} 
\begin{figure}
\centering 
\includegraphics[width=0.479\textwidth]{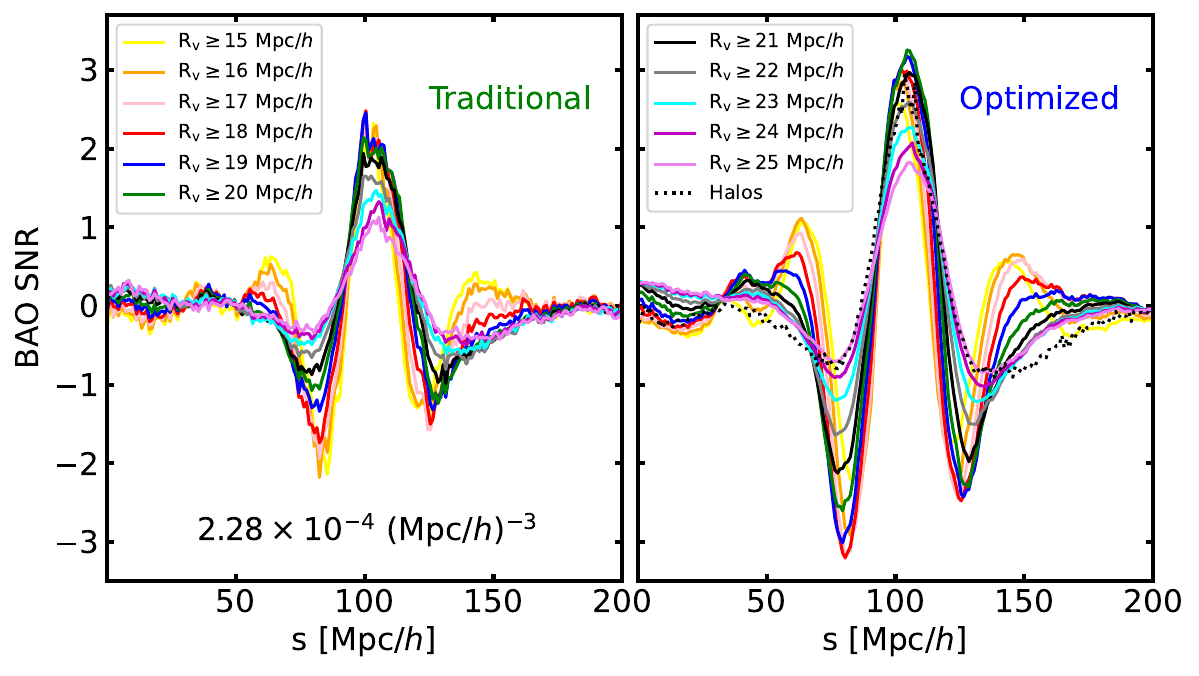}
\caption{The BAO SNRs of large DT voids with various radius cutoffs (solid lines), obtained from traditional (left panel) and optimized (right panel) schemes. For comparison, the result from corresponding halo catalogues with number density of $\bar{n}_\mathrm{h} = 2.28 \times 10^{-4} \, (h^{-1} \mathrm{Mpc})^{-3}$ is also shown as a dashed line in the right panel. A key finding is that, under optimized scheme, large DT voids with an optimal cutoff can yield even stronger SNRs than their corresponding halos, highlighting their potentials as more effective tracers of early-Universe signatures such as BAOs (cf. Figure~\ref{fig:large_DT_PS_BAO}).}
\label{fig:large_DT_halos_SNR}
\end{figure}

Void statistics refer to statistical analyses of the distribution and properties of low-density regions, providing unique insights into LSS formation and evolution. As such, complementary statistical information to other summary statistics can be leveraged to break degeneracies and tighten cosmological constraints (\citealt{2022A&A...661A.146B, 2022ApJ...935..100K, 2023MNRAS.522..152P, 2025ApJ...993..227V, 2025arXiv250408221S, 2025PhRvD.112f3516S, 2026arXiv260114362C}). Moreover, a combination of void statistics and other summary statistics may allow for a better systematic calibration to enhance the robustness of cosmological measurements, as systematics affect each statistic differently (\citealt{2021JCAP...01..028Z}). Particularly, void dynamics remain near-linear even in present epoch (\citealt{2021MNRAS.500.4173S, 2023JCAP...05..031S, 2025arXiv250907092S}). This potentially renders underdense regions cleaner probes than overdense regions for extracting cosmological signatures encoded in pristine LSS, e.g., BAO signals measurable through void clustering (see Figure~\ref{fig:large_DT_halos_SNR} and Figure~\ref{fig:large_DT_PS_BAO}). 

In this section, we thoroughly evaluate the performance of our proposed optimized measurement scheme (see Section~\ref{sec:optimization}) by comparing the tessellation-based void statistics obtained with traditional and optimized schemes, focusing on three most extensively studied metrics, i.e., VSF, VTCF, and VPS. In this study, cosmic voids are identified in halo distributions with different mass cutoffs, corresponding to different halo number densities (see Section~\ref{sec:data}). In essence, these voids are not necessarily representative of those in the underlying mass distribution, as different matter tracers can produce distinct structural patterns (\citealt{2015MNRAS.454..889N, 2019MNRAS.488.3526C, 2024MNRAS.529.4325B}).

\begin{figure*}
\centering 
\includegraphics[width=0.99\textwidth]{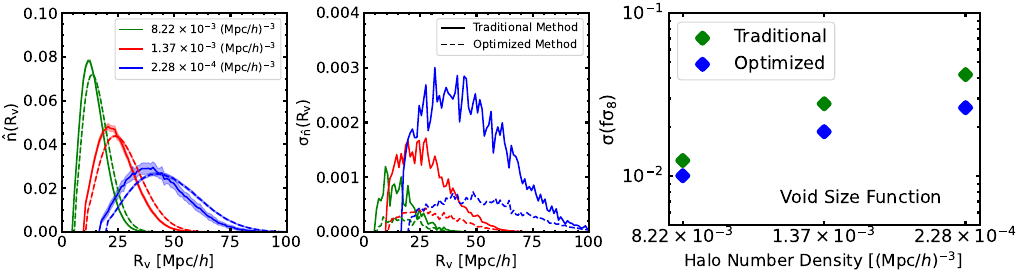}
\caption{VSFs of ZOBOV/VIDE voids (left panel), their $1$-$\sigma$ scatters (middle panel), and Fisher error forecasts for $f\sigma_8$ (right panel). Green diamonds: traditional method; blue diamonds: optimized method (the same below). As shown, compared to traditional scheme, optimized scheme reduces the statistical errors of VSFs by a factor of $3\text{--}4$ and tightens the $f\sigma_8$ constraints by $19\text{--}37\%$, with larger gains at lower halo number densities.}
\label{fig:VIDE_VSF_Fisher}
\end{figure*}  

\subsection{Void Size Function} \label{subsec:VSF}
The use of VSF has been predicated on the theoretical analysis of hierarchically evolving void population by \citealt{2004MNRAS.350..517S}, who derived an analytical expression for a "peaked" void size distribution, assuming voids to be spherical in shape. Based on the excursion set formalism and the identification of void-in-void and void-in-cloud processes, the resulting size distribution explicitly depends on matter power spectrum and hence on cosmology, and was later refined by \citealt{2012MNRAS.420.1648P} and \citealt{2013MNRAS.434.2167J}. Subsequently, VSF was demonstrably shown to be sensitive to the underlying cosmological model (\citealt{2015PhRvD..92h3531P, 2015JCAP...11..018M}), further establishing the basis for its use in cosmological analyses. Building upon those early foundations, VSF has recently emerged as a promising tool for constraining cosmology in modern galaxy surveys (\citealt{2019MNRAS.488.3526C, 2022A&A...667A.162C, 2023ApJ...953...46C, 2024MNRAS.532.1049S, 2025MNRAS.tmp..898S, 2025A&A...695A..19F, 2025ApJ...993..227V}).

This statistic characterizes the abundance of voids as a function of their size. For spherical DT void, size is directly characterized by its radius, while for irregularly shaped ZOBOV/VIDE void, size is described by its effective radius, i.e., the radius of sphere with same volume as the void. If a ZOBOV/VIDE void consists of $n$ Voronoi cells, its effective radius is determined collectively by all these constituent cells,
\begin{equation}\label{eq:3}
R_{\mathrm{eff}}=\left(\frac{3}{4\pi}\sum_{i=1}^{n}V_{\mathrm{cell}}^{i}\right)^{1/3},
\end{equation}
where $V_{\mathrm{cell}}^{i}$ represents the volume of the $i$-th Voronoi cell belonging to the void (see Section~\ref{sec:voids}).

In the literature, VSF is typically quantified with void number density, $\bar{n}_\mathrm{v}$, as a function of void radius, $R_\mathrm{v}$, i.e., $\hat{n}(R_\mathrm{v}) = \mathrm{d}\bar{n}_\mathrm{v}/\mathrm{d}R_\mathrm{v}$. In this work, while adopting a conceptually similar framework, we employ normalized void number function (\citealt{2016MNRAS.459.2670Z}) as a proxy for VSF:
\begin{equation}\label{eq:4}
\hat{n}_\mathrm{v}\left(R_{\mathrm{v}}\right)=\frac{\mathrm{d}N_\mathrm{v}/\mathrm{d}R_\mathrm{v}}{N_{\mathrm{total}}},
\end{equation}
which satisfies normalization condition,
\begin{equation}\label{eq:5}
\int\hat{n}_\mathrm{v}\left(R_{\mathrm{v}}\right)\mathrm{d}R_\mathrm{v}=1,
\end{equation}
where $N_\mathrm{total}$ is the total void number of the sample\footnote{Note that, by definition, $\hat{n}_\mathrm{v}(R_\mathrm{v})$ is exactly the probability density function for void radii, representing the likelihood of finding a void of radius $R_\mathrm{v}$.}. In practice, we estimate $\hat{n}_\mathrm{v}(R_{\mathrm{v}})$ by binning voids into radius bins $\left[R_{\mathrm{v}}^i-\Delta R_{\mathrm{v}}/2,R_{\mathrm{v}}^i+\Delta R_{\mathrm{v}}/2\right]$ and computing
\begin{equation}\label{eq:6}
\hat{n}_{\mathrm{v}}\left(R_{\mathrm{v}}^i\right)\simeq\frac{\Delta N_{\mathrm{v}}\left(R_{\mathrm{v}}^i\right)}{N_{\text{total}}\cdot\Delta R_\mathrm{v}},
\end{equation}
where $\Delta N_{\mathrm{v}}\left(R_{\mathrm{v}}^i\right)$ denotes the number of voids in the $i$-th bin and bin width is fixed at $\Delta R_{\mathrm{v}} = 1\,h^{-1}\mathrm{Mpc}$.

In Figure~\ref{fig:DT_all_disjoint_VSF}, we present the VSFs of all DT voids and disjoint DT voids, measured with traditional and optimized methods, respectively. To highlight the improvements of our optimized scheme over traditional one, we also plot corresponding statistical uncertainties separately in the same figure. Here, these uncertainties are calculated as the standard deviations from 40 realizations of our BAO-featured simulations, except for the highest halo number density case, where 10 realizations are used (see Section~\ref{sec:data}; the same below). 

We observe a systematic shift of VSFs toward larger void radii under optimized scheme, relative to traditional one. This is a natural result, as optimized scheme involves a subsampling procedure (see Figure~\ref{fig:flow_char}), causing the identified voids to originate from halo catalogues with effectively lower number densities, and thus be systematically larger. As expected, we see that VSFs measured with optimized scheme exhibit smaller uncertainties than those obtained from traditional scheme. Specifically, for all DT voids with dense number densities, the uncertainties are reduced by about $1/4\text{--}1/3$, with stronger improvements for void samples of relatively lower number densities. For more sparse disjoint DT voids, the reductions are even more dramatic, reaching nearly a factor of $4\text{--}5$! These results provide preliminary evidences for the effectiveness of our optimized scheme in significantly reducing statistical uncertainties of Delaunay-based void statistics.

Encouragingly, our optimization approach is effective not only for Delaunay-based DT voids but also for Voronoi-based ZOBOV/VIDE voids, which are more commonly adopted in cosmological applications. This is demonstrated in Figure~\ref{fig:VIDE_VSF_Fisher}, which shows the results for ZOBOV/VIDE voids with sparse number densities. To assess the gains in cosmological constraining power from optimized scheme over traditional one, we also present Fisher error forecasts for structure growth rate $f\sigma_8$ (see Appendix~\ref{app:Fisher_forecast}), a key cosmological parameter widely constrained in the analyses of redshift-space distortions (RSDs).

We find that optimized scheme can reduce VSF uncertainties of ZOBOV/VIDE voids by a factor of about $3\text{--}4$ compared to the case of traditional scheme. These reductions translate into $\sim 19\text{--}37\%$ improvements in constraining power on $f\sigma_8$, with advantages becoming more pronounced at lower halo number densities (cf.\ Figure~\ref{fig:VIDE_PS_Fisher} for the case of VPS). These results demonstrate that optimized scheme is particularly effective for low-number-density types of voids, probably because tessellation-based statistics become increasingly unstable in such regimes, whereas our subsampling-and-averaging strategy can effectively suppress these instabilities and thereby greatly reduce their statistical uncertainties. Moreover, it suggests that, for typically-sparse voids, the statistical errors of void statistics are dominated by uncertainties arising from tessellation instabilities rather than from unavoidable cosmic variances. These conclusions likewise hold for other void statistics, as shown below.

\begin{figure*}
\centering 
\includegraphics[width=0.99\textwidth]{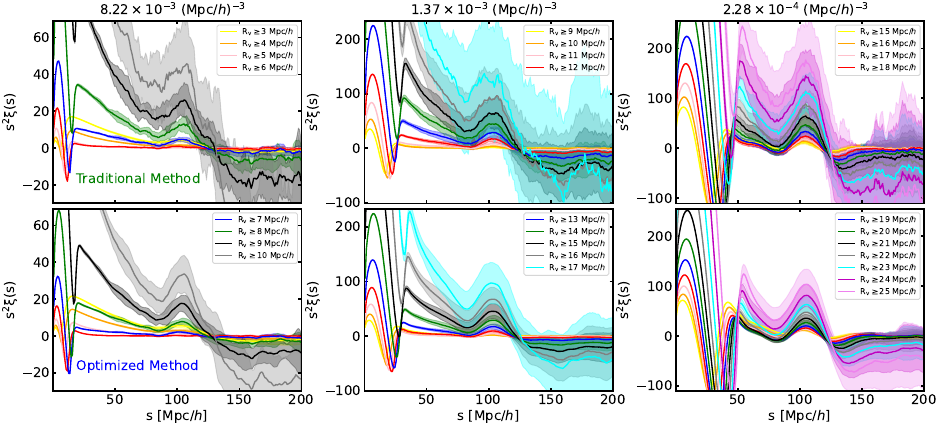}
\caption{VTCFs of large DT voids measured with traditional scheme (upper panels) and optimized scheme (bottom panels). Results are shown for large DT voids with various radius cutoffs, as specified in the legends. The three columns correspond to our halo catalogues with different number densities, from left to right: $\bar{n}_\mathrm{h} = 8.22 \times 10^{-3} \, (h^{-1} \mathrm{Mpc})^{-3}$, $\bar{n}_\mathrm{h} = 1.37 \times 10^{-3} \, (h^{-1} \mathrm{Mpc})^{-3}$, and $\bar{n}_\mathrm{h} = 2.28 \times 10^{-4} \, (h^{-1} \mathrm{Mpc})^{-3}$, respectively. Compared to the case of traditional scheme, the VTCFs obtained from optimized scheme exhibit smoother trends and smaller uncertainties. These improvements lead to enhanced BAO SNRs (see Figure~\ref{fig:large_DT_halos_SNR}), facilitating more accurate determinations of BAO scales.}
\label{fig:large_DT_TCF}
\end{figure*}

\begin{figure*}
\centering 
\includegraphics[width=\textwidth]{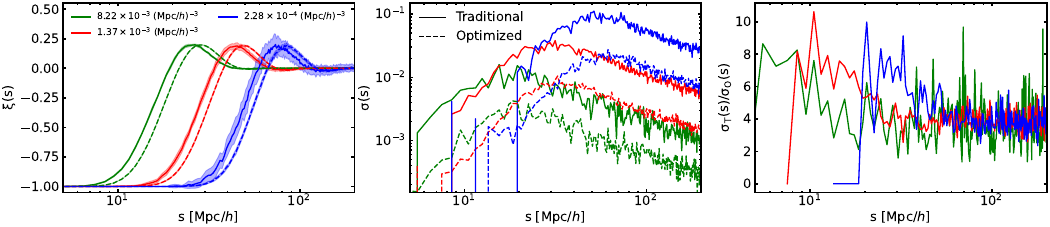}
\caption{VTCFs of ZOBOV/VIDE voids (left panel) and their statistical errors (middle panel) under traditional (solid lines) and optimized (dashed lines) schemes, together with the error ratios between two measurement schemes (right panel). This demonstrates that optimized scheme yields VTCFs with statistical uncertainties roughly a factor of $4$ lower than those from traditional scheme, highlighting its substantial advantage.}
\label{fig:VIDE_TCF}
\end{figure*}

\subsection{Void Two-point Correlation Function} \label{subsec:VTCF}
VTCF of large DT voids has been routinely employed to perform BAO measurements (e.g., \citealt{2016PhRvL.116q1301K, 2016MNRAS.459.4020L, 2020MNRAS.491.4554Z, 2022MNRAS.511.5492Z, 2023MNRAS.526.2889T, 2023MNRAS.521.4731V}), while VTCF of ZOBOV/VIDE voids has also been widely applied in various cosmological studies (e.g., \citealt{2014JCAP...12..013H, 2015JCAP...11..018M, 2019MNRAS.488.4413K, 2022ApJ...935..100K}). Additionally, void-galaxy cross-correlation function of ZOBOV/VIDE-like voids was extensively employed to constrain cosmology (e.g., \citealt{2017JCAP...07..014H, 2019PhRvD.100b3504N, 2020MNRAS.499.4140N, 2020JCAP...06..012H, 2020JCAP...12..023H, 2022MNRAS.516.4307W, 2022A&A...658A..20H, 2023MNRAS.523.6360W, 2023A&A...677A..78R, 2025ApJ...993..227V, 2025JCAP...06..001F}). Since this work focuses on voids alone, we defer a systematic exploration of such cross-analyses (or combined-analyses, e.g., \citealt{2020MNRAS.491.4554Z, 2022MNRAS.511.5492Z}) to future studies.

VTCF quantifies the excess probability, relative to a random distribution, of finding a pair of voids separated by a given distance $s$. Its Fourier counterpart is VPS (see Section~\ref{subsec:VPS}), and both characterize the clustering properties of void spatial distribution. To measure these statistics, void positions are required to be specified, and this depends on how voids are physically identified. For DT voids, positions are simply their sphere centers, whereas for ZOBOV/VIDE voids, they are defined as volume-weighted barycenters\footnote{A dynamical definition identifies void positions as expansion centers from which streamlines emanate. In practice, these expansion centers do not necessarily coincide with void barycenters, which are strongly influenced by surrounding structures. Hence, the extent to which the two-point correlation function can provide stringent cosmological constraints remains under discussion, given the uncertainties in the choice of void centers.},
\begin{equation}\label{eq:7}
\mathbf{X}_{\mathrm{v}}=\frac{\sum_{i=1}^{n}\mathbf{x}_{i}V_{\text{cell}}^{i}}{\sum_{i=1}^{n}V_{\text{cell}}^{i}},
\end{equation}
where $\mathbf{x}_{i}$ indicates the position of the $i$-th halo/galaxy in the void (see Section~\ref{sec:voids}). In this work, the Peebles–Hauser estimator of two-point correlation function (TCF) (\citealt{1974ApJS...28...19P}) is adopted to obtain an unbiased VTCF, 
\begin{equation}\label{eq:8}
\xi_\mathrm{vv}(s)=\frac{\text{DD}(s)}{\text{RR}(s)}-1,
\end{equation}
where $\text{DD}(s)$ and $\text{RR}(s)$ are normalised pair counts within separation bin $\left[s-\Delta s/2,s+\Delta s/2\right]$ from data and random catalogues, respectively. This formula takes advantage of utilization of data from simulation boxes with periodic boundary conditions and free from observational systematics. In such a scenario, $\text{RR}(s)$ term can be computed analytically through
\begin{equation}\label{eq:9}
\text{RR}(s)=\frac{4\pi}{3}\frac{(s+\Delta s/2)^3-(s-\Delta s/2)^3}{V},
\end{equation}
where $V$ is the volume of simulation box and bin width is set to $\Delta s=1\,h^{-1}\mathrm{Mpc}$. For implementation, a publicly available {\tt FCFC} code\footnote{\url{https://github.com/cheng-zhao/FCFC}} (\citealt{2023A&A...672A..83Z}) is employed for calculations.

In Figure~\ref{fig:large_DT_TCF}, we present the VTCFs of large DT voids with various radius cutoffs, obtained from traditional and optimized schemes. These voids are constructed from our halo catalogues with BAO features (see Section~\ref{sec:data}). Consistent with expectations, the results show that the VTCFs under optimized scheme are noticeably smoother and exhibit smaller scatter than those from traditional scheme, especially on BAO scales, clearly demonstrating the superior performance of our optimization method for this statistic.

In particular, we further take advantage of the paired simulations with and without BAO features to estimate the BAO SNR\footnote{Alternatively, one can compute the difference between TCFs of tracers from each pair of simulations with and without BAO features, and define the BAO SNR as $\overline{\Delta\xi}_\mathrm{wig-now}/\sigma_\mathrm{wig-now}$, where $\overline{\Delta\xi}_\mathrm{wig-now}$ and $\sigma_\mathrm{wig-now}$ denote the mean and corresponding standard deviation across multiple such differences. In fact, this definition will result in amplitudes larger than those from Equation~\ref{eq:10} due to sample-variance cancellations, but it does not alter our conclusions.} as
\begin{equation}\label{eq:10}
\text{BAO\ SNR} = \frac{\bar{\xi}_\mathrm{wig}(s)-\bar{\xi}_\mathrm{now}(s)}{\sqrt{\sigma_\mathrm{wig}^2(s)+\sigma_\mathrm{now}^2(s)}},
\end{equation}
where $\bar{\xi}_\mathrm{wig}(s)$ and $\bar{\xi}_\mathrm{now}(s)$ denote the mean TCFs of tracers obtained by averaging over multiple realizations of simulations with and without BAO features, respectively, and $\sigma_\text{wig}(s)$ and $\sigma_{\text{now}}(s)$ are their corresponding standard deviations. 

The results for large DT voids with different radius cutoffs are presented in Figure~\ref{fig:large_DT_halos_SNR}. Here, these voids are created from the halo catalogues of $\bar{n}_\mathrm{h} = 2.28 \times 10^{-4}\,(h^{-1}\mathrm{Mpc})^{-3}$, which roughly corresponds to the typical number density of BOSS CMASS luminous red galaxies (\citealt{2011AJ....142...72E, 2013AJ....145...10D}). We see that BAO SNR of large DT voids varies with radius cutoff, attaining its maximum at an optimal radius cutoff of about $17$ and $19\,h^{-1}\mathrm{Mpc}$ for traditional and optimized schemes, respectively. This trend has been reported in previous studies (e.g., \citealt{2016MNRAS.459.4020L, 2022MNRAS.511.5492Z}), while our determination of optimal cutoff is more rigorous. Particularly, owing to substantial reduction in VTCF uncertainties (see Figure~\ref{fig:large_DT_TCF}), the BAO SNRs obtained under optimized scheme are significantly enhanced compared to those derived from traditional scheme. This improvement is crucial, as it allows for a more accurate determination of BAO characteristic scale (i.e., the BAO peak position), which will be elaborated in our forthcoming papers.

\begin{figure*}
\centering 
\includegraphics[width=0.98\textwidth]{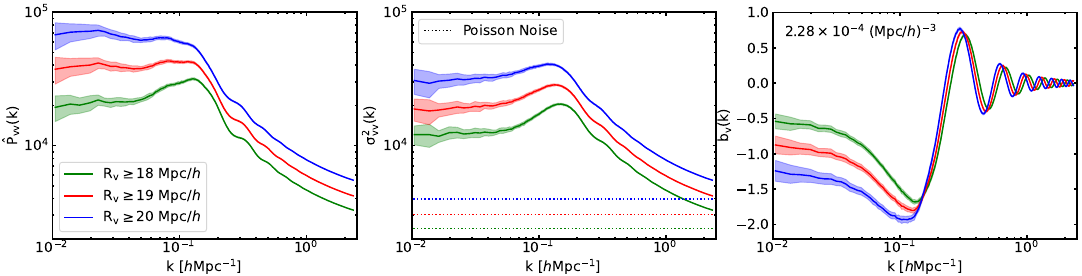}
\caption{Directly-measured VPSs (left panel), shot noises (middle panel), and biases (right panel) for large DT voids with radius cutoffs of $18$, $19$, and $20\,h^{-1}\mathrm{Mpc}$. Here, voids are constructed from our halo catalogues with number density of $\bar{n}_\mathrm{h} = 2.28 \times 10^{-4}\,(h^{-1} \mathrm{Mpc})^{-3}$ and analyzed under traditional scheme. For reference, the middle panel also shows Poisson predictions for corresponding shot noises.}
\label{fig:large_DT_PS_SN_bias}
\end{figure*} 

\begin{figure}
\centering 
\includegraphics[width=0.47\textwidth]{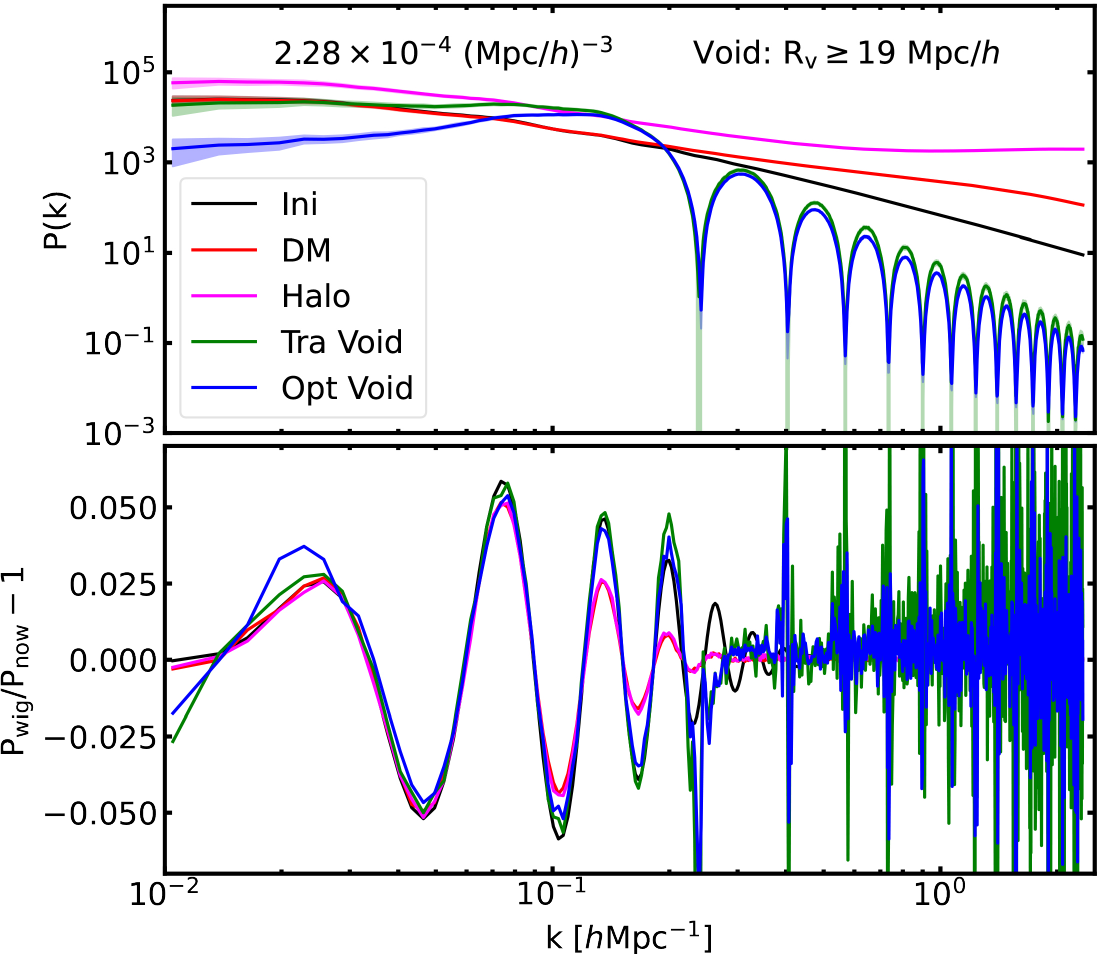}
\caption{Shot-noise-subtracted VPSs for large DT voids with radius cutoff of $19\,h^{-1}\mathrm{Mpc}$ under traditional and optimized schemes (upper panel), and their corresponding normalized BAO signatures (lower panel). For comparison, we also show the results for corresponding halo catalogues with $\bar{n}_\mathrm{h} = 2.28 \times 10^{-4}\,(h^{-1}\mathrm{Mpc})^{-3}$, as well as for dark matter and initial conditions. As demonstrated, the amplitudes of normalized BAO signatures from dark matter and halos are significantly damped by LSS non-linearities, especially on small scales. In contrast, regardless of measurement scheme, the BAO signatures extracted from large DT voids align well with the linear theory prediction. These results suggest that density troughs are less affected by LSS non-linear dynamics, which appears to help preserve the linear BAO signatures (cf.\ Figure~\ref{fig:large_DT_halos_SNR}).}
\label{fig:large_DT_PS_BAO}
\end{figure}

For reference, the BAO SNR of corresponding halos is also displayed in Figure~\ref{fig:large_DT_halos_SNR}, enabling a direct comparison with the conventional result. As shown, under traditional scheme, halo BAO SNR is more pronounced than that of voids with optimal radius cutoff (cf.\ \citealt{2020MNRAS.491.4554Z, 2022MNRAS.511.5492Z}). Interestingly, however, under optimized scheme, the optimal void BAO SNR even surpasses that of halos, benefiting from substantial scatter suppressions. This novel phenomenon may be intuitively explained by the fact that these density troughs are less affected by LSS non-linear gravitational dynamics (\citealt{2021MNRAS.500.4173S, 2023JCAP...05..031S, 2025arXiv250907092S}), allowing them to better preserve the linear BAO signatures (cf.\ Figure~\ref{fig:large_DT_PS_BAO} and \citealt{2014PhRvL.112d1304H}). This intriguing hypothesis can be further rigorously tested by exploiting dedicated cosmological simulation data sets, which warrants detailed investigations in future studies\footnote{In our ongoing work, we further employ extensive sets of MultiDark-Patchy mock galaxy catalogues (\citealt{2016MNRAS.456.4156K}) to constrain the BAO characteristic scales, using VTCFs of large DT voids with different radius cutoffs.}.

In Figure~\ref{fig:VIDE_TCF}, we present VTCFs of ZOBOV/VIDE voids\footnote{The results for disjoint DT voids are similar to those for ZOBOV/VIDE voids, as both are non-overlapping void populations with relatively low number densities. For brevity, we do not show them further here.}, which exhibit bumps at void exclusion scales, corresponding to roughly twice the mean void radii (\citealt{2014JCAP...12..013H, 2014arXiv1409.7621H, 2019MNRAS.488.4413K, 2022ApJ...935..100K}). This feature, also visible in VTCFs of large DT voids (see the bumps on the left side of BAO peaks in Figure~\ref{fig:large_DT_TCF}), marks the characteristic scales at which voids are most tightly packed. Naturally, ZOBOV/VIDE voids obtained with optimized scheme display larger exclusion scales compared to those yielded from traditional scheme, as they are correspondingly larger. Within exclusion scales, VTCFs of ZOBOV/VIDE voids gradually decline toward $-1$ as void separation decreases (\citealt{2015JCAP...11..018M}), because of mutual void exclusion\footnote{It refers to the fact that ZOBOV/VIDE voids do not overlap and therefore cannot be arbitrarily close to one another; see also \citealt{2013PhRvD..88h3507B, 2021PhRvD.103h3530B} for similar effects in TCFs of halos.}. Similar effects are also evident in VTCFs of large DT voids (see Figure~\ref{fig:large_DT_TCF}), corresponding to the dips on the left side of exclusion bumps. However, in such a case, VTCFs rise again to be positive at smaller void separations, reflecting their overlapping nature. Beyond exclusion scales, the clustering of ZOBOV/VIDE voids progressively diminishes, and no discernible BAO feature is observed on BAO scales. Importantly, regardless of the case, optimized scheme can significantly reduce VTCF uncertainties of ZOBOV/VIDE voids by a factor of $\sim 4$ compared with traditional scheme, representing substantial improvements.

\subsection{Void Power Spectrum} \label{subsec:VPS} 
\begin{figure*}
\centering 
\includegraphics[width=0.999\textwidth]{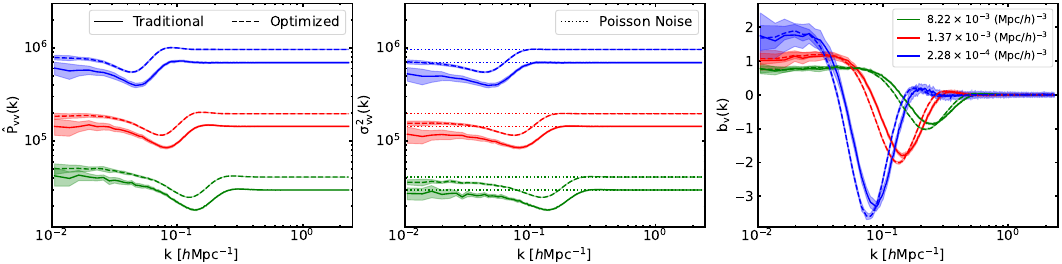}
\caption{Directly-measured VPSs (left panel), shot noises (middle panel), and biases (right panel) for ZOBOV/VIDE voids obtained under traditional (solid lines) and optimized (dashed lines) schemes. For comparison, Poisson predictions for corresponding shot noises are also displayed in the middle panel.}
\label{fig:VIDE_PS_SN_bias}
\end{figure*}

VPS, denoted as $P_\mathrm{vv}(k)$, also serves as a widely used statistical tool for extracting cosmological information from LSS (\citealt{2014JCAP...12..013H, 2014PhRvD..90j3521C, 2019JCAP...12..055S, 2019MNRAS.488.4413K, 2024ApJ...976..244S}). To measure this statistic, we construct void fields onto $N_\mathrm{g}^3=1024^3$ grids using Triangular-Shaped Cloud (TSC) mass assignment method, which can help mitigate the aliasing effects caused by finite mesh sampling (\citealt{2016MNRAS.460.3624S}). In Fourier domain, window function effects introduced by TSC interpolation are corrected through the expression (\citealt{2005ApJ...620..559J})
\begin{equation}\label{eq:11}
W_\mathrm{TSC}(\mathbf{k})=\prod_{i\in\{x,y,z\}}\left[\frac{\sin \left(\pi k_i/2k_{\mathrm{Nyq}}\right)}{\pi k_i/2k_{\mathrm{Nyq}}}\right]^3,
\end{equation}
where $k_i$ denotes the component of wavevector $\mathbf{k}$ along Cartesian axis $i=x,y,z$ with total magnitude $k=\|\mathbf{k}\|=(k_{x}^{2}+k_{y}^{2}+k_{z}^{2})^{1/2}$, and $k_{\mathrm{Nyq}}=\pi N_{\mathrm{g}}/L_{\mathrm{box}}$ is the Nyquist frequency. Ultimately, to estimate the VPSs, we adopt linear binning scheme in $k$-space with a fixed width of $\mathrm{d}k=0.003\,h\mathrm{Mpc}^{-1}$, spanning wavenumbers from $k_\mathrm{min}=0.009\,h\mathrm{Mpc}^{-1}$ up to $k_{\mathrm{Nyq}}=2.331\,h\mathrm{Mpc}^{-1}$. For practical calculations, we employ the functionality provided by publicly available {\tt nbodykit} package \citep{2018AJ....156..160H}.

Similar to halos (\citealt{2009PhRvL.103i1303S, 2010PhRvD..82d3515H, 2021ApJS..254....4L}), voids also act as biased and stochastic tracers of underlying dark matter. In Fourier space, the relation between void overdensity $\delta_\mathrm{v}({\bf k})$ and dark matter overdensity $\delta_\mathrm{m}({\bf k})$ can be expressed as
\begin{equation}\label{eq:14}
\delta_\mathrm{v}({\bf k}) = b_\mathrm{v}(k)\delta_\mathrm{m}({\bf k}) + \varepsilon_\mathrm{v}({\bf k}),
\end{equation}
where void bias (\citealt{2014PhRvL.112d1304H, 2014PhRvD..90j3521C, 2019JCAP...12..055S, 2020ApJ...889...89C}) is defined as
\begin{equation}\label{eq:15}
b_\mathrm{v}(k)=\dfrac{\langle\delta_\mathrm{v}({\bf k})\delta_\mathrm{m}({\bf k})^*\rangle}{\langle\vert\delta_\mathrm{m}({\bf k})\vert^2\rangle}
\end{equation}
and stochastic term $\varepsilon_\mathrm{v}({\bf k})$ is assumed to be uncorrelated with dark matter field (i.e., $\langle\varepsilon_\mathrm{v}({\bf k})\delta_\mathrm{m}({\bf k})^*\rangle = 0$). Hence, the directly-measured VPS, $\hat{P}_\mathrm{vv}(k)$, can be decomposed into two terms
\begin{equation}\label{eq:16}
\hat{P}_\mathrm{vv}(k) = P_\mathrm{vv}(k) + \sigma_\mathrm{vv}^2(k)
\end{equation}
with
\begin{equation}\label{eq:17}
P_\mathrm{vv}(k) = b_\mathrm{v}^2(k)P_\mathrm{mm}(k),
\end{equation}
where $\sigma_\mathrm{vv}^2(k) = \langle\vert\varepsilon_\mathrm{v}({\bf k})\vert^2\rangle$ is void shot noise term, $P_\mathrm{mm}(k) = \langle\vert\delta_\mathrm{m}({\bf k})\vert^2\rangle$ is dark matter power spectrum, and $P_\mathrm{vv}(k)$ denotes the true VPS without shot-noise contribution (cf.\ \citealt{2019MNRAS.488.4413K}). This decomposition implies that $\sigma_\mathrm{vv}^2(k)$ encapsulates all sources of stochasticity between voids and dark matter (\citealt{2009PhRvL.103i1303S}). For uniformly-weighted void fields, $\sigma_\mathrm{vv}^2(k)$ is usually naively modelled as Poisson shot noise, which is given by the inverse of void number density $1/\bar{n}_\mathrm{v}$ (\citealt{2014PhRvL.112d1304H, 2014PhRvD..90j3521C, 2019JCAP...12..055S}).

The directly-measured VPSs of large DT voids with radius cutoffs of $18$, $19$, and $20\,h^{-1}\mathrm{Mpc}$, obtained via traditional scheme, are shown in the left panel of Figure~\ref{fig:large_DT_PS_SN_bias}. Here, voids are constructed from our halo catalogues with number density of $\bar{n}_\mathrm{h} = 2.28 \times 10^{-4}\,(h^{-1}\mathrm{Mpc})^{-3}$ (see Section~\ref{sec:data}). To elucidate the behaviours of shot noises and biases for large DT voids, the middle panel of Figure~\ref{fig:large_DT_PS_SN_bias} illustrates corresponding shot-noise contributions (see Equation~\ref{eq:16}) alongside their Poisson predictions for reference, while the right panel of the same figure depicts the associated void biases (see Equation~\ref{eq:15}). Given that the results obtained from optimized scheme exhibit similar overall trends to those yielded by traditional scheme, yet feature markedly reduced statistical uncertainties, the relevant discussions are omitted for conciseness.

\begin{figure*}
\centering 
\includegraphics[width=0.999\textwidth]{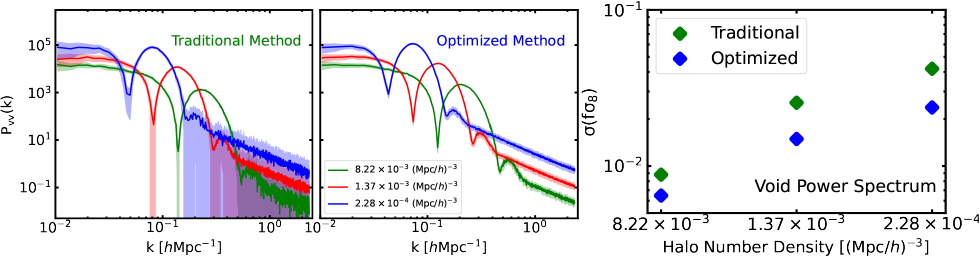}
\caption{Shot-noise-subtracted VPSs for ZOBOV/VIDE voids obtained under traditional (left panel) and optimized (middle panel) schemes (cf.\ Figure~\ref{fig:VIDE_PS_SN_bias}), and Fisher error forecasts for $f\sigma_8$ (right panel). As expected, compared to traditional scheme, optimized scheme substantially reduces the statistical errors of VPSs and tightens the $f\sigma_8$ constraints by $26$–$43\%$, with larger gains at lower halo number densities.}
\label{fig:VIDE_PS_Fisher}
\end{figure*} 

Shot noise is typically non-Poissonian due to tracer clustering and exclusion effects, which give rise to super-Poissonian and sub-Poissonian shot noise, respectively (see\ \citealt{2009PhRvL.103i1303S, 2013PhRvD..88h3507B, 2017MNRAS.470.2566P, 2017PhRvD..96h3528G, 2021ApJS..254....4L} and Refs. therein). Although large DT voids exhibit some degree of exclusion, their mutual overlap causes clustering to dominate over exclusion effects on small scales (see\ Figure~\ref{fig:large_DT_TCF}). As a result, their shot noises are super-Poissonian on large scales and gradually approach Poissonian level on smaller scales. These unwanted extra components can significantly affect the measurements of VPSs and therefore should be properly accounted for and removed (see\ Figure~\ref{fig:large_DT_PS_BAO}). We also find that their biases are negative on large scales, as these void-in-void-type voids strictly trace underdense regions and therefore exhibit a negative correlation with underlying dark matter (see\ Equation~\ref{eq:15} and Figure~\ref{fig:cross-correlation_coefficients}). In addition, these void biases exhibit dips around void exclusion scales, $k_\mathrm{exc} \sim \pi/\bar{R}_\mathrm{v}$, where $\bar{R}_\mathrm{v}$ is the mean void radius (see\ \citealt{2014PhRvL.112d1304H} and Figure~\ref{fig:VIDE_PS_SN_bias} for similar behaviours). As $k$ increases, they oscillate around zero with gradually decreasing amplitudes and eventually asymptote to zero in the limit $k \to +\infty$ (cf.\ \citealt{2016MNRAS.459.2670Z}). Both of these features can be attributed to void exclusion effects (\citealt{2014PhRvL.112d1304H}). In Appendix~\ref{app:VPS_bias_SN}, we further explore the results under traditional scheme for all DT voids and disjoint DT voids; interested readers are referred there for details. 

In particular, the upper panel of Figure~\ref{fig:large_DT_PS_BAO} presents the shot-noise-subtracted VPSs (see Equation~\ref{eq:17}) for large DT voids with radius cutoff of $19\,h^{-1}\mathrm{Mpc}$, measured under traditional and optimized schemes. Here, for reference, we also show shot-noise-subtracted halo power spectrum, together with linear and non-linear dark matter power spectra. Moreover, the bottom panel of Figure~\ref{fig:large_DT_PS_BAO} shows the normalized BAO signatures for different tracers, computed as $P_\mathrm{wig}(k)/P_\mathrm{now}(k)-1$, where $P_\mathrm{wig}(k)$ and $P_\mathrm{now}(k)$ are the shot-noise-subtracted power spectra of tracers from simulations with and without BAO features, respectively (see Section~\ref{sec:data}).

As illustrated, the shot-noise-subtracted VPSs of large DT voids exhibit prominent oscillations as $k$ increases, consistent with the behaviours of corresponding void biases (see Figure~\ref{fig:large_DT_PS_SN_bias} and Equation~\ref{eq:17}). Here, each spike-like dip corresponds to a zero-crossing of associated void bias, a pattern also observed in the shot-noise-subtracted VPSs of other void types (see Figure~\ref{fig:VIDE_PS_Fisher} and Figure~\ref{fig:DT_disjoint_PS}). Traditional and optimized schemes yield qualitatively similar results, whereas optimized scheme appreciably reduces the statistical scatters. As is well known, the BAO signatures derived from halos and dark matter are considerably damped on relatively small scales due to bulk flows arising from LSS non-linear evolution (\citealt{2007ApJ...664..660E, 2012MNRAS.427.2132P}), in contrast to ideal linear BAOs. Remarkably, we find that void BAOs preserve linear BAO amplitudes quite well\footnote{Noted that this striking finding relies on proper removals of the highly influential shot-noise components (cf.\ \citealt{2020MNRAS.491.4554Z} for the results with shot noises), which requires the knowledge of underlying dark matter field (see Equation~\ref{eq:17}), inaccessible in galaxy redshift surveys. Nevertheless, assuming no correlation between void shot noise and halo shot noise, and that halo shot noise is Poissonian, one may roughly estimate shot-noise-free VPSs via
\begin{equation}\label{eq:18}
P_\mathrm{vv}(k) \approx \frac{P^2_\mathrm{vh}(k)}{\hat{P}_\mathrm{hh}(k)-1/\bar{n}_\mathrm{h}},
\end{equation}
where $P_\mathrm{vh}(k)$ is the void-halo cross power spectrum (cf.\ \citealt{2019JCAP...12..055S}). However, this lies beyond the scope of the present study and will be explored in detail in future work.}, seemingly suggesting that density troughs retain early-Universe information more effectively than density peaks (cf.\ \citealt{2014PhRvL.112d1304H, 2021MNRAS.500.4173S, 2023JCAP...05..031S, 2025arXiv250907092S}). It is worth emphasizing that this interesting finding leverages the inherent benefits of large DT voids, which naturally pinpoint void-in-void local density minima at centers of empty Delaunay spheres with radii exceeding a radius threshold (see Section~\ref{sec:voids}), eliminating the necessity for explicit density field reconstruction employed in methods like spherical-underdensity algorithms (e.g., \citealt{2005MNRAS.363..977P, 2015MNRAS.451.1036C}). Importantly, this pragmatic approach ensures a sizable tracer sample\footnote{Potentially, one could construct analogous tracers by structure classification algorithms (e.g., \citealt{2022A&A...661A.146B, 2025PhRvD.112f3516S}) or by simply selecting positions of density minima below a specified density threshold in a smoothed field (cf.\ \citealt{2005MNRAS.363..977P}). These methodologies might yield results comparable to ours. Determining which of these strategies is optimal for maximizing BAO SNRs, however, is left for future work.}, leading to decreased shot noise, improved statistical robustness, and consequently reinforcing the significance of void BAO signatures (cf.\ \citealt{2021PhRvD.103d3502C} for a comparison with other BAO results associated with underdense regions).

In the left panel of Figure~\ref{fig:VIDE_PS_SN_bias}, we demonstrate the directly-measured VPSs for ZOBOV/VIDE voids constructed from our halo catalogues with three different number densities, comparing the results obtained with traditional and optimized schemes (cf.\ \citealt{2014PhRvL.112d1304H, 2015A&C.....9....1S, 2019JCAP...12..055S}). In addition, the same figure presents their corresponding void shot noises, together with related Poisson predictions, in the middle panel, and their associated void biases in the right panel.

Due to the sparsity of ZOBOV/VIDE voids, their directly-measured VPSs are dominated by shot noises, which drive the plateaus observed on intermediate and small scales. By construction, ZOBOV/VIDE voids are non-overlapping (\citealt{2015A&C.....9....1S}); therefore, they exhibit exceptionally strong exclusion, dominating over their clustering effects. As a result, their shot noises turn out to be sub-Poissonian on large scales, but approach Poissonian predictions on sufficiently small scales (cf.\ \citealt{2014PhRvL.112d1304H})\footnote{When subtracting shot noises components from the directly-measured VPSs of ZOBOV/VIDE voids, naively modeling them as purely Poissonian would result in negative shot-noise-subtracted VPSs on large scales (cf.\ \citealt{2014PhRvD..90j3521C}), which is manifestly unphysical, as power spectra must be positive definite.}. Moreover, we observe that the void biases remain nearly scale-independent (cf.\ \citealt{2014PhRvL.112d1304H, 2014PhRvD..90j3521C, 2019JCAP...12..055S, 2020ApJ...889...89C}) and positive on large scales. This is attributable to the fact that the ZOBOV/VIDE voids adopted here contain a non-negligible fraction of void-in-cloud objects, whose spatial distribution moderately follows that of dark matter, resulting in positive correlations between dark matter and ZOBOV/VIDE voids (cf.\ Equation~\ref{eq:15}, Figure~\ref{fig:cross-correlation_coefficients}, and \citealt{2014PhRvL.112d1304H, 2014PhRvD..90j3521C, 2020ApJ...889...89C}). Similar to large DT voids (see Figure~\ref{fig:large_DT_PS_SN_bias}), the biases of ZOBOV/VIDE voids also exhibit pronounced negative dips near void exclusion scales, resulting from void exclusion, which induces anti-correlations between voids and underlying dark matter on those scales (cf.\ \citealt{2014PhRvL.112d1304H, 2019JCAP...12..055S, 2020ApJ...889...89C}). On small scales, the void biases converge toward zero, reflecting negligible correlation between dark matter and ZOBOV/VIDE voids in that regime (cf.\ \citealt{2014PhRvL.112d1304H, 2014PhRvD..90j3521C}). Overall, the incorporation of a sampling process in the optimized method leads to sparser void subsamples, thereby increasing the shot-noise contributions in the directly-measured VPSs. Nonetheless, optimized scheme significantly reduces the statistical uncertainties across all results relative to traditional scheme.

In Figure~\ref{fig:VIDE_PS_Fisher}, we proceed to display the shot-noise-subtracted VPSs of ZOBOV/VIDE voids measured under both schemes (cf.\ \citealt{2019MNRAS.488.4413K}), along with their corresponding Fisher error forecasts for $f\sigma_8$. It is found that these VPSs exhibit distinct bumps around void exclusion scales, corresponding to the pronounced dips in ZOBOV/VIDE void biases on the same scales (cf.\ Figure~\ref{fig:VIDE_PS_SN_bias} and Equation~\ref{eq:17}). Crucially, as anticipated, the optimized scheme substantially suppresses the statistical errors of the VPSs. Accordingly, the constraining power of the VPSs on $f\sigma_8$ is greatly enhanced under optimized scheme relative to traditional scheme, with an improvement of $\sim 26\text{--}43\%$. These results further highlight the compelling application potentials of our optimization method for future galaxy-survey data analyses.

\section{Summaries and Discussions} \label{sec:summary}
The sensitivities of tessellation constructions to perturbations in input points often results in numerical instabilities for tessellation-based algorithms (cf.\ \citealt{2021MNRAS.503..557A}), leading to notable additional statistical uncertainties in the derived statistics. In general, especially for scenarios of voids with relatively low number densities, theses uncertainties overwhelmingly dominate the total statistical noises over the cosmic variances, thereby substantially downgrading the cosmological constraining power of tessellation-based statistics.

In this work, we propose an optimization method based on subsampling and averaging to address this critical issue inherent in tessellation-based statistics. As a case study, we implement our new approach on two representative tessellation-based void finders, namely Delaunay-based DIVE (\citealt{2016MNRAS.459.2670Z}) and Voronoi-based ZOBOV/VIDE (\citealt{2015A&C.....9....1S}), to investigate three standard void statistics, i.e., VSF, VTCF, and VPS. We find that our optimized scheme can greatly suppresses the excess statistical noise induced by tessellation instabilities and substantially enhance the cosmological constraining power of void statistics. Specifically, for the VSF and VPS of ZOBOV/VIDE voids, compared to traditional scheme, optimized scheme improves the constraints on $f\sigma_8$ by $\sim 19\text{--}37\%$ and $\sim 26\text{--}43\%$, respectively, depending on halo number densities, with lower-density samples exhibiting larger improvements. These gains are equivalent to an enlargement of effective survey volume by $\sim 52\text{--}152\%$ for VSF and $\sim 83\text{--}200\%$ for VPS under traditional scheme.

Interestingly, we find that, under optimized scheme, the BAO SNR obtained from VTCF of large DT voids can even outperform that of corresponding halos (see Figure~\ref{fig:large_DT_halos_SNR}). This suggests that density troughs may serve as more effective tracers than density peaks for extracting cosmological information imprinted on the pristine LSS of early Universe, e.g., BAO signatures. Furthermore, under both traditional and optimized schemes, we observe that after appropriately removing shot-noise contributions, the amplitudes of normalized BAO signatures extracted from VPSs of large DT voids can be quite close to that of ideal linear BAO signature (see Figure~\ref{fig:large_DT_PS_BAO}). This appears to indicate that density troughs are less affected by LSS non-linear dynamics, thereby mitigating the BAO damping inevitably suffered by halos/galaxies, and provides further support for our preliminary conclusions. These encouraging findings, reported for the first time in the literature, point toward more detailed and systematic future explorations of density troughs for extracting key cosmological signatures of early Universe, e.g., BAO features, primordial non-Gaussianities, etc.

The optimized measurement scheme for tessellation-based statistics advocated in this work is straightforward to implement, requiring no complex algorithm design, while offering powerful error reduction for various tessellation-based void statistics. This would make our method particularly well-suited for wide adoption in future galaxy-survey data analyses. Beyond that, our new scheme can not only reduce statistical uncertainties but also be extended to simultaneously correct for various observational systematics, which otherwise introduce additional anisotropy and inhomogeneity into the galaxy distributions observed from redshift surveys. This can be achieved by inverting survey systematic effects during subsampling, where galaxies are assigned probabilities based on their weights and survey selection functions (see \citealt{2025RASTI...4...14N})\footnote{It lies beyond the scope of the present work, and we will explore this direction in detail in future studies.}. Here, although generating multiple subsamples increases computational costs compared to traditional scheme, it has a crucial advantage of avoiding reliance on computationally expensive and labour intensive forward modeling of these systematics for parameter inference (see \citealt{2025RASTI...4...14N}). Building on this rationale, our optimized scheme holds promise as a standard framework for tessellation-based statistics in future galaxy surveys.

It is also worth noting that while the subsampling in our optimized scheme reduces the effective halo/galaxy number densities and introduces systematic shifts in resulting statistics compared to traditional scheme, it does not fundamentally change their analytical modeling. This is because cosmological statistics are inherently measured from a halo/galaxy sample with a specific number density, and the results from optimized scheme merely correspond to a scenario of halo/galaxy catalogue with a lower number density. In any case, emulators for these statistics can always be constructed from simulations and directly integrated with the optimized measurement scheme, which poses no methodological obstacle. Moreover, while our analyses concentrate on tessellation-based void statistics, the proposed new scheme should be broadly applicable to other tessellation-based statistics, e.g., DTFE Minkowski functionals (\citealt{2024ApJS..273...33L}) and Voronoi volume function (\citealt{2020MNRAS.495.3233P}). For instance, we notice that it appears capable of reducing the statistical uncertainties of Voronoi volume function (see \citealt{2025arXiv250616408D}).

\section*{Acknowledgments}
Y.L.\ would like to thank Pengjie Zhang, Yuting Wang, Dandan Xu, Jiajun Zhang, Shengyu He, Chen Su, Anning Gao, Kwan Chuen Chan, Yi Zheng, Jiaxi Yu, Leander Thiele, Jia Liu, Ji Yao, and Charling Tao for useful communications. We also thank the anonymous referee for helpful comments, which helped make the paper more informative. This work was supported by National Natural Science Foundation of China (Nos.\ 12303005, 12273015, 12503112, 12433003), Swiss National Science Foundation research grant ``Cosmology with 3D Maps of the Universe'' with Nos.\ 200020\_207379 and 10006415, National Key R\&D Program of China (Nos.\ 2023YFA1605600, 2023YFA1607800, 2023YFA1607802), and science research grant from China Manned Space Project with No.\ CMS-CSST-2025-A04. Y.L.\ acknowledges the support from Shuimu Tsinghua Scholar Program (No.\ 2022SM173). Y.Y. and Z.D. acknowledge the sponsorship from Yangyang Development Fund. This work made use of the Gravity Supercomputer at the Department of Astronomy, Shanghai Jiao Tong University.

\appendix
\begin{figure*}
\centering 
\includegraphics[width=0.921\textwidth]{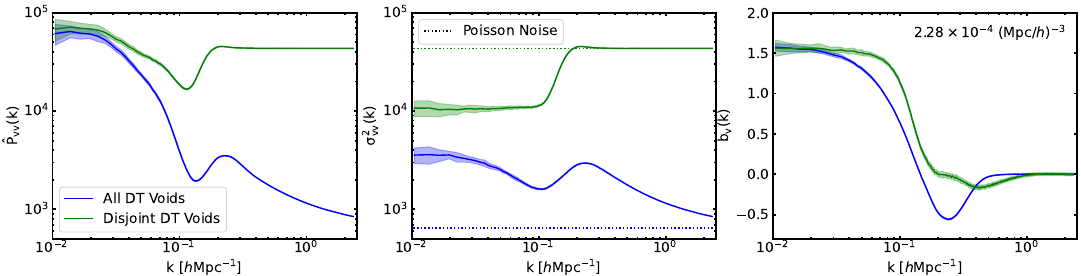}
\caption{Directly-measured VPSs (left panel), shot noises (middle panel), and biases (right panel) for all DT voids (blue lines) and disjoint DT voids (green lines). These results are measured under traditional scheme. Here, voids are constructed from our halo catalogues with number density of $\bar{n}_\mathrm{h} = 2.28 \times 10^{-4}\,(h^{-1} \mathrm{Mpc})^{-3}$. For reference, Poisson predictions for corresponding shot noises are also shown in the middle panel.}
\label{fig:DT_disjoint_PS_SN_bias}
\end{figure*} 

\begin{figure}
\centering 
\includegraphics[width=0.47\textwidth]{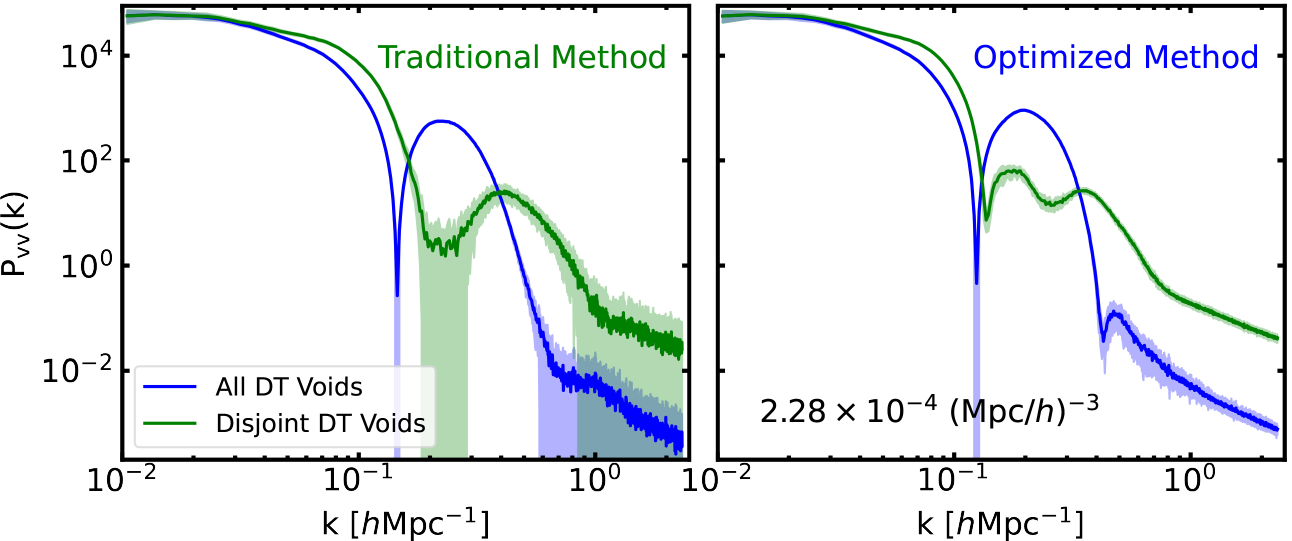}
\caption{Shot-noise-subtracted VPSs for all DT voids and disjoint DT voids, measured under traditional (left panel) and optimized (right panel) schemes (cf.\ Figure~\ref{fig:DT_disjoint_PS_SN_bias}). As before, optimized scheme achieves significant uncertainty suppressions in VPSs relative to traditional scheme, further demonstrating its substantial advantage.}
\label{fig:DT_disjoint_PS}
\end{figure}

\section{Bootstrap Subsampling} \label{app:subsampling}
For a dataset containing $n$ samples, bootstrap resampling with replacement is performed to generate a new set of $n$ samples. For any given sample $x$ in the original dataset, the probability of it not being selected in a single draw is $1-1/n$. After $n$ independent draws with replacement, the probability that it is never selected is $(1-1/n)^n$, and the probability that it is selected at least once is $1-(1-1/n)^n$. Therefore, over many bootstrap repetitions, on average, approximately $1-(1-1/n)^n$ of the original samples will be included in each resampled set, while about $(1-1/n)^n$ of the samples will remain unselected. From the perspective of limit convergence, as $n \to +\infty$, $(1-1/n)^n$ tends toward $1/e$, which is $\sim 36.8\%$. Thus, the proportion of samples that are selected converges to $\sim 63.2\%$. In our subsampling method, the samples selected at least once in resampling form a subset of the original dataset, comprising $\sim 63.2\%$ of the original samples.

\section{Tessellation-Based Density Field Estimators} \label{app:Field_Estimator}
The blocky, piecewise-constant density fields constructed by zeroth-order Voronoi Tessellation Field Estimator (VTFE), which assigns to each point a density proportional to the inverse of its corresponding Voronoi cell volume, exhibits sharp discontinuities at cell boundaries, leading to pronounced artifacts inherent to Voronoi tessellation structure. The Delaunay Tessellation Field Estimator (DTFE; \citealt{1996MNRAS.279..693B, 2000A&A...363L..29S, 2009LNP...665..291V}), in its first-order formulation, alleviates some of these issues. It estimates density at each tracer position by inverting volume of its contiguous Voronoi cell (i.e., the union of all Delaunay tetrahedra sharing that tracer as a vertex), followed by a multi-dimensional linear interpolation within each Delaunay tetrahedron, yielding a continuous field. However, it still exhibits prominent artifacts, typically appearing as triangular imprints of the DTFE kernels, particularly in poorly sampled regions.

Nevertheless, these two density field estimators form the basis of watershed-transform void-finding techniques. The first watershed void finder, WVF, was introduced by \citealt{2007MNRAS.380..551P}, based on a first-order DTFE density field. Subsequently, ZOBOV (cf.\ Section~\ref{sec:voids}), introduced by \citealt{2008MNRAS.386.2101N}, instead employs zeroth-order VTFE density estimates combined with a more elaborate statistical treatment and forms the core of various void-finding pipelines, such as VIDE (\citealt{2015A&C.....9....1S}), REVOLVER (\citealt{2019PhRvD.100b3504N}), and V$^2$ (\citealt{2022JOSS....7.4033D, 2023ApJS..265....7D}). Moreover, DTFE is also applied in various advanced pipelines for characterizing, identifying, and classifying structures of LSS. Examples include MMF (\citealt{2007A&A...474..315A}) and NEXUS (\citealt{2013MNRAS.429.1286C}), which leverage multi-scale geometry of structural components, and SpineWeb (\citealt{2010ApJ...723..364A}) and DisPerSE (\citealt{2011MNRAS.414..350S, 2011MNRAS.414..384S}), which utilize topology of cosmic web via Morse theory (\citealt{morse1934calculus, milnor1963morse, jost2008riemannian}).

\begin{figure*}
\centering 
\includegraphics[width=0.95\textwidth]{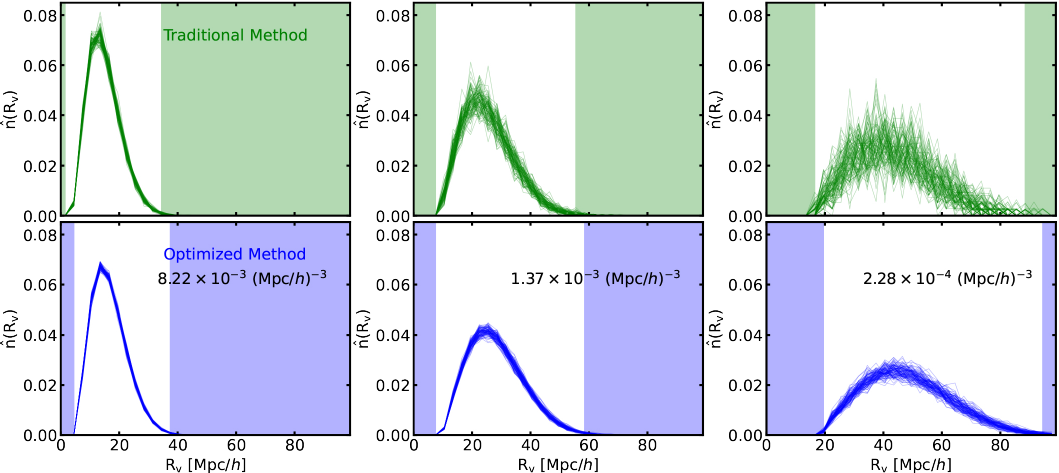}
\caption{VSFs of smaller-volume subsamples of ZOBOV/VIDE voids in redshift space. Upper panels show the results obtained with traditional scheme, while lower panels present those obtained from optimized scheme. From left to right, the panels correspond to halo number densities of $\bar{n}_\mathrm{h} = 8.22 \times 10^{-3} \, (h^{-1} \mathrm{Mpc})^{-3}$, $\bar{n}_\mathrm{h} = 1.37 \times 10^{-3} \, (h^{-1} \mathrm{Mpc})^{-3}$, and $\bar{n}_\mathrm{h} = 2.28 \times 10^{-4} \, (h^{-1} \mathrm{Mpc})^{-3}$, respectively. For each panel, we present 100 randomly selected VPSs from 270 smaller-volume subsamples for the case of $\bar{n}_\mathrm{h} = 8.22 \times 10^{-3} \, (h^{-1} \mathrm{Mpc})^{-3}$ and from 1080 smaller-volume subsamples for the cases of $\bar{n}_\mathrm{h} = 1.37 \times 10^{-3} \, (h^{-1} \mathrm{Mpc})^{-3}$ and $\bar{n}_\mathrm{h} = 2.28 \times 10^{-4} \, (h^{-1} \mathrm{Mpc})^{-3}$ (see Appendix~\ref{app:Fisher_forecast}). Here, the unshielded regions denote the radius ranges actually used in our Fisher error forecasts for $f\sigma_8$ (see Table~\ref{tab:info}).}
\label{fig:VIDE_VSF_range}
\end{figure*}

\begin{table*}
\resizebox{0.96\linewidth}{!}{
\begin{tabular}{lccc}
\hline \hline \begin{tabular}{l}
\text{$[R_\mathrm{lower},R_\mathrm{upper}]$}
\end{tabular} & \begin{tabular}{c}
\text{$\bar{n}_\mathrm{h} = 8.22 \times 10^{-3} \, (h^{-1} \mathrm{Mpc})^{-3}$}
\end{tabular} & \begin{tabular}{c}
\text{$\bar{n}_\mathrm{h} = 1.37 \times 10^{-3} \, (h^{-1} \mathrm{Mpc})^{-3}$}
\end{tabular} & \begin{tabular}{c}
\text{$\bar{n}_\mathrm{h} = 2.28 \times 10^{-4} \, (h^{-1} \mathrm{Mpc})^{-3}$}
\end{tabular} \\
\hline \begin{tabular}{l}
\textbf{Traditional VSFs} 
\end{tabular} & \begin{tabular}{c}
$[1.5,34.5],34$
\end{tabular} & \begin{tabular}{c}
$[7.5,55.5],49$ 
\end{tabular} & \begin{tabular}{c}
$[16.5,88.5],73$
\end{tabular} \\ \begin{tabular}{l}
\textbf{Optimized VSFs} 
\end{tabular} & \begin{tabular}{c}
$[4.5,37.5],34$ 
\end{tabular} & \begin{tabular}{c}
$[7.5,58.5],52$ 
\end{tabular} & \begin{tabular}{c}
$[19.5,94.5],76$ \\
\end{tabular} \\
\hline \hline
\end{tabular}
}
\caption{Radius ranges and numbers of data points of VSFs of ZOBOV/VIDE voids used in our Fisher error forecasts for $f\sigma_8$. For both traditional and optimized schemes, we show these values in various halo number density cases (cf.\ Figure~\ref{fig:VIDE_VSF_range}).}
\label{tab:info}
\end{table*}

\begin{figure}
\centering 
\includegraphics[width=0.43\textwidth]{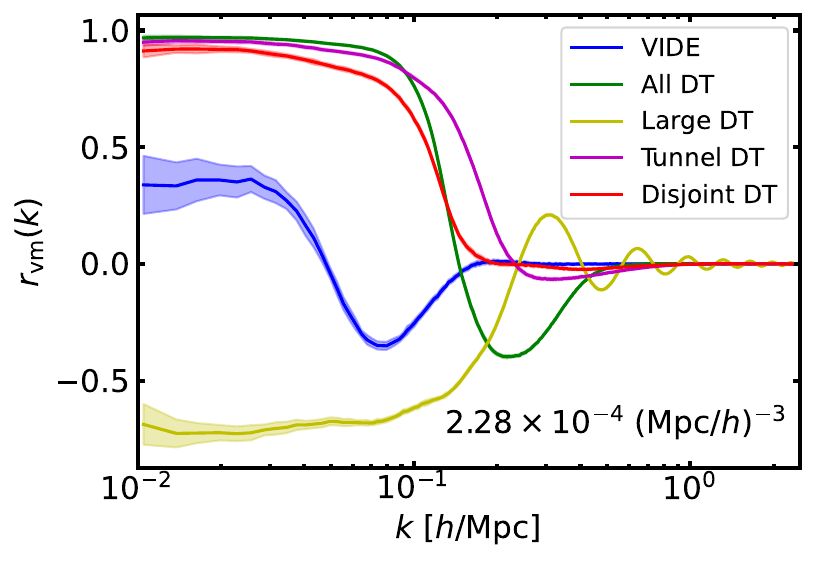}
\caption{Cross-correlation coefficients between different types of voids and underlying dark matter. The results for ZOBOV/VIDE, all DT, large DT, tunnel DT, and disjoint DT voids are indicated by blue, green, yellow, magenta, and red lines, respectively. Here, these void populations are constructed from our halo catalogues with number density of $\bar{n}_\mathrm{h} = 2.28 \times 10^{-4} \, (h^{-1} \mathrm{Mpc})^{-3}$, and large DT voids are defined as those with radius cutoff of $19\,h^{-1}\mathrm{Mpc}$.}
\label{fig:cross-correlation_coefficients}
\end{figure}

\section{VPSs, Biases, and Shot Noises of All DT Voids and Disjoint DT Voids} \label{app:VPS_bias_SN}
For the scenario of halo catalogues with $\bar{n}_\mathrm{h} = 2.28 \times 10^{-4}\,(h^{-1}\,\mathrm{Mpc})^{-3}$, we present in the left panel of Figure~\ref{fig:DT_disjoint_PS_SN_bias} the directly-measured VPSs of both all DT voids and disjoint DT voids, obtained with traditional scheme. Their corresponding shot-noise components are shown in the middle panel of Figure~\ref{fig:DT_disjoint_PS_SN_bias}, together with their Poisson predictions as references, while the associated void biases are displayed in the right panel of the same figure.

As discussed in Section~\ref{subsec:VPS}, the strong overlap among all DT voids causes their clustering to overwhelm void exclusion, resulting in super-Poissonian shot noises on large scales. In contrast, for disjoint DT voids, void exclusion decisively dominates over clustering, leading to sub-Poissonian shot noises on large scales. Nevertheless, on small scales, the shot noises of both samples approach their respective Poisson predictions. Similar to other types of voids (cf.\ Section~\ref{subsec:VPS}), the corresponding void biases are basically scale-independent on large scales, exhibit pronounced dips around exclusion scales, and tend toward zero on small scales.

In Figure~\ref{fig:DT_disjoint_PS}, we further show the shot-noise-subtracted VPSs of all DT voids and disjoint DT voids, measured under traditional and optimized schemes in the left and right panels, respectively. As in the case of ZOBOV/VIDE voids, distinct bumps are also visible around void exclusion scales, corresponding to the negative dips of their associated void biases on the same scales, induced by void exclusion effects. In particular, as anticipated, compared with traditional scheme, optimized scheme yields significantly smaller statistical uncertainties in these noise-free VPSs.

\section{Fisher Error Forecasts for Structure Growth Rate} \label{app:Fisher_forecast} 
In our Fisher error forecasts for $f\sigma_8$, we adopt the method developed in \citealt{2024ApJS..273...33L}. This method relies on analyses performed in redshift space by rescaling halo velocities. For brevity, we omit a detailed description of the procedure here and refer interested readers to the Appendix C of \citealt{2024ApJS..273...33L} for full details. In this appendix, we provide only necessary explanations for certain technical points.

To mitigate the issues associated with non-invertibility of covariance matrices, the radius range [$R_\mathrm{lower}$, $R_\mathrm{upper}$] adopted for VSFs of ZOBOV/VIDE voids is progressively broadened with decreasing halo number density, thereby increasing the number of data points included in the analyses. This treatment is illustrated in Figure~\ref{fig:VIDE_VSF_range} (see the unshaded regions) and summarized in Table~\ref{tab:info}. For VPSs of ZOBOV/VIDE voids, we employ $28$ uniformly spaced data points over the wavenumber interval $k \in [0.048, 0.855]$ in the Fisher error forecasts.

To effectively increase the number of realizations for computing covariance matrices, each void sample in redshift space\footnote{Note that, in our optimized method, this refers to one void subsample constructed from a downsampled halo catalogue in redshift space (see Section~\ref{sec:optimization} and Figure~\ref{fig:flow_char}).} is divided into $n_f = 3^3 = 27$ smaller-volume subsamples through an equal spatial partitioning of the simulation box. Thus, for a given void statistic, we obtain in total $n_t = 10 \times n_f = 270$ realizations for $\bar{n}_\mathrm{h} = 8.22 \times 10^{-3} \, (h^{-1} \mathrm{Mpc})^{-3}$, and $n_t = 40 \times n_f = 1080$ realizations for $\bar{n}_\mathrm{h} = 1.37 \times 10^{-3} \, (h^{-1} \mathrm{Mpc})^{-3}$ and $\bar{n}_\mathrm{h} = 2.28 \times 10^{-4} \, (h^{-1} \mathrm{Mpc})^{-3}$. Given the numbers of data points used, these realizations are sufficient to ensure robust covariance estimates.

\section{Cross-correlation Coefficients Between Voids and Dark Matter} \label{app:cross-correlation}
As a supplement, in Figure~\ref{fig:cross-correlation_coefficients}, under traditional scheme, we present the cross-correlation coefficients between different types of voids and dark matter, defined as $r_\mathrm{vm}(k) \equiv P_\mathrm{vm}(k)/\sqrt{\hat{P}_\mathrm{vv}(k)\, \hat{P}_\mathrm{mm}(k)}$, where $P_\mathrm{vm}(k)$ is the cross-power spectrum of voids and dark matter, and $\hat{P}_\mathrm{vv}(k)$ and $\hat{P}_\mathrm{mm}(k)$ are their respective auto-power spectra. We find that large DT voids are negatively correlated with dark matter on large scales, as they are strictly selected to be void-in-voids (see Figure~\ref{fig:DT_voids_halos} and Section~\ref{sec:voids}). In contrast, other classes of voids exhibit positive correlations with dark matter on large scales, because these void populations are not filtered and thus include a fraction of voids-in-clouds tracing high-density regions (cf.\ Figure~\ref{fig:DT_voids_halos}). Nevertheless, they display negative dips on their respective void exclusion scales due to void exclusion. Furthermore, for all void types, their correlations with dark matter gradually approach zero as $k \to +\infty$.


\bibliographystyle{yahapj}
\bibliography{void}

\end{document}